\documentclass[preprint,12pt]{article}

\usepackage{bbm}
\usepackage{amsmath}
\usepackage{amssymb}
\usepackage{stmaryrd}
\usepackage{graphicx}
\usepackage{amsthm}
\usepackage{caption}
\usepackage{accents}

\newcommand*{\bdot}[1]{\accentset{\mbox{\large\bfseries .}}{#1}}

\title{On rate-dependent dissipation effects in electro-elasticity}
\author{Prashant Saxena, Duc Khoi Vu, Paul Steinmann \\
Chair of Applied Mechanics, University of Erlangen--Nuremberg, \\ Egerlandstra\ss e 5, 91058 Erlangen, Germany}
\date{}

\begin{document}

\maketitle

\begin{abstract}
 This paper deals with the mathematical modelling of large strain electro-viscoelastic deformations in electro-active polymers. Energy dissipation is assumed to occur due to mechanical viscoelasticity of the polymer as well as due to time-dependent effective polarisation of the material. Additive decomposition of the electric field $\mathbb{E} = \mathbb{E}_e + \mathbb{E}_v$ and  multiplicative decomposition of the deformation gradient $\mathbf{F} = \mathbf{F}_e \mathbf{F}_v$ are proposed to model the internal dissipation mechanisms. The theory is illustrated with some numerical examples in the end. 
\end{abstract}


\\

Original version published in the \emph{International Journal of Nonlinear Mechanics} 62: 1--11 (2014) doi: 10.1016/j.ijnonlinmec.2014.02.002

\section{Introduction}

Over the past few decades, the theory and numerical simulation of the coupled electro-mechanical problem have been interesting subjects of research, cf. Pao \cite{Pao1978} and Eringen and Maugin \cite{Eringen1990}.  However, with the invention of the so-called  electro-active polymers (EAPs) capable of exhibiting large deformations in response to the application of electric fields, several new challenges appear and need to be addressed. Open problems remain both in the understanding of electro-mechanical coupling in soft matter and in simulating the behavior of electro-sensitive bodies under the influence of an electric field.

EAPs can be used as alternatives to materials traditionally used to develop actuators like piezoelectric ceramics, shape memory metals and electro-rheological fluids, cf. O'Halloran et al.~\cite{O'Halloran2008}.  Potential applications of EAPs in developing artificial muscles and robotic systems include robot manipulators \cite{Wingert2006}, soft pumps \cite{Bowers2011}, loud-speakers \cite{Heydt2008}, portable force feed-back devices \cite{Zhang2006}, haptic interfaces \cite{Ozsecen2010}, electric generators for energy harvesting \cite{Pelrine2001}--\cite{McKay2011}, transport vehicles \cite{Anderson2011}, \cite{Michel2008}, and sensing equipment \cite{Toth2002}--\cite{Son2009}, among others.

Efforts were made in the past to model and simulate the behaviour of EAPs using the theory of nonlinear elasticity and nonlinear visco-elasticity, for example, by  Kofod \cite{Kofod2001}, Sommer-Larsen et al.~\cite{Sommer-Larsen2002}, Goulbourne et al.~\cite{Goulbourne2003}, Yang et al.~\cite{Yang2004, Yang2006}, and Rosset et al.~\cite{Rosset2008}. However, the papers mentioned above assume that the material electric properties are independent of deformation. Note that because large strain occurs during the deformation process, the nonlinearity of the material electric properties must be accounted for. In order to overcome this shortcoming, some related boundary-value problems involving finite deformation were analyzed by taking into account the nonlinearity of the electric polarisation, for example, in the works of Voltairas et al. \cite{Voltairas2003}, Dorfmann and Ogden \cite{Dorfmann2005}, M\"uller et al.~\cite{Mueller2010}, Zwecker et al.~\cite{Zwecker2011}, and Vertechy et al.~\cite{Vertechy2012}. The effect of viscosity in the modeling of EAPs was examined recently by Ask et al.~\cite{Ask2012} and B\"uschel et al.~\cite{Bueschel2013}.

A basic assumption in the modelling of EAPs in the papers mentioned above has been an instantaneous or `elastic' response of the material to an applied electric field. This, however, may not be the case in all the electroactive polymers and we aim at modelling this phenomena in this research. We work under a more general case where it is assumed that on the application of an electric field, the overall macroscopic polarisation of the material is time-dependent. Thus, in addition to the mechanical viscoelasticity of the polymeric matrix, an additional energy-dissipating mechanism is considered on account of the evolution of the electric polarisation with time.

Among the several approaches towards a phenomenological theory of mechanical viscoelasticity, the literature is usually divided on account of the nature of internal variable used to quantify dissipation. The internal variable can be assumed to be of stress-type, as proposed by Simo \cite{Simo1987} and Lion \cite{Lion1997}, or it can be strain-type, as used by Lubliner \cite{Lubliner1985}, Reese and Govindjee \cite{Reese1998} and Huber and Tsakmakis \cite{Huber2000}. In the latter approach, which has been also followed in this paper, the deformation gradient is decomposed into elastic and inelastic parts ($\mathbf{F} = \mathbf{F}_e \mathbf{F}_v$)  where the inelastic part is determined from a differential type flow rule. In addition to the mechanical viscoelastic dissipation, we also model electric dissipation by considering a similar decomposition of our independent variable (electric field in this case) into `elastic' and `viscous' parts as $\mathbb{E} = \mathbb{E}_e + \mathbb{E}_v$. This follows a similar approach by Saxena et al.~\cite{Saxena2013a} for the magnetic counterpart of this problem.
The energy and momentum balance laws of electroelasticity are derived from the fundamental equations of electrostatics following the work of McMeeking and Landis \cite{McMeeking2005}.

This paper is organised as follows.
The theory of rate-dependent electroelastic deformations is presented in Section 2. Starting with the basic principles  of electrostatics and continuum mechanics (as detailed in the Appendix), we obtain the energy and momentum balance laws in the case of electroelasticity. The deformation gradient and the electric field are decomposed into equilibrium and non-equilibrium parts ($\mathbf{F} = \mathbf{F}_e \mathbf{F}_v, \mathbb{E} = \mathbb{E}_e + \mathbb{E}_v$). Using the laws of thermodynamics and a form of the free energy density function, constitutive equations are derived along with the conditions to be satisfied by the evolution equations of the non-equilibrium quantities.

For the purpose of obtaining numerical solutions later, the energy density function and the evolution equations for the non-equilbrium quantities are specialised to specific forms. Several electro-visco-elastic coupling parameters are introduced  in this step and we define  thermodynamically consistent and physically reasonable evolution laws for the internal variables. In Section 3, numerical solutions are obtained corresponding to five different types of (mechanical and electric) loading conditions. The effects of the underlying deformation,  strain rate, electric field, and electric field rate are studied on the evolution of the resulting stress and the dielectric displacement. The results, presented graphically, show a strong coupling between strain and electric field, as well as the strong dependence of the response on electro-viscoelastic coupling parameters thus making the model amenable to fitting with experimental data, as an when it becomes available in future.

\section{Theory}
We consider an electroelastic material that, when undeformed,  unstressed and in the absence of electric fields, occupies the material configuration $\mathcal B_0$ with boundary $\partial \mathcal B_0$. It is then subjected to a static deformation due to the combined action of an electric field, mechanical surface tractions and body forces.
The spatial configuration at time $t$ is denoted by $\mathcal B_t$ with a boundary $\partial \mathcal B_t$. The two configurations are related by a deformation function $\boldsymbol \chi$ which maps every point $\mathbf X \in \mathcal B_0$ to a point $\mathbf x = \boldsymbol \chi(\mathbf X, t) \in \mathcal B_t$.
The deformation gradient is defined as $\mathbf{F} = \mbox{Grad}\, \boldsymbol \chi$, where Grad is the gradient operator with respect to $\mathbf{X}$. Its determinant is given by $J = \mbox{det} \, \mathbf{F}$.

\subsection{Balance laws and boundary conditions}

\subsubsection{Equations of electrostatics}

Let $q$ be the electric charge density per unit volume in $\mathcal B_t$, $\mathbbm{e}$ be the spatial electric field vector, $\mathbbm{d}$ be the spatial electric displacement vector, and $\mathbbm{p}$ be the spatial polarisation vector.
The balance equations for the electric quantities are given by a simplified form of the two Maxwell's equations as
\begin{equation} \label{gov euler 1}
\mbox{curl}\, \mathbbm{e} = \mathbf{0}, \quad \mbox{div}\, \mathbbm{d} = q,
\end{equation}
where the electric vectors are related by the constitutive law
\begin{equation}
\mathbbm{d} = \varepsilon_0 \mathbbm{e} + \mathbbm{p},
\end{equation}
and curl and div denote the corresponding differentiation with respect to the position vectors $\mathbf{x}$ in the spatial configuration $\mathcal B_t$.

We note  that the above equations can also be written in the material configuration $\mathcal B_0$ by employing the following transformations
\begin{equation} \label{pullback}
 \mathbb{E} = \mathbf{F}^t \mathbbm{e}, \quad \mathbb{D} = J \mathbf{F}^{-1} \mathbbm{d}, \quad \mathbb{P} = J \mathbf{F}^{-1} \mathbbm{p},
\end{equation}
thus giving
\begin{equation}
 \mbox{Curl}\, \mathbb{E} = 0, \quad \mbox{Div}\, \mathbb{D} = J q, \quad \mathbb{D} = \varepsilon_0 J \mathbf{C}^{-1} \mathbb{E} + \mathbb{P},
\end{equation}
such that Curl and Div denote the corresponding differentiation operators with respect to the position vectors $\mathbf{X}$ in $\mathcal B_0$ and $\mathbf{C} = \mathbf{F}^t \mathbf{F}$ is the right Cauchy--Green deformation tensor.

Since curl of a gradient vanishes, the electric field vector can be written as the gradient of a scalar potential from equation \eqref{gov euler 1}$_1$ as
\begin{equation} \label{E potential}
\mathbbm{e} = - \mbox{grad}\, \phi.
\end{equation}

At an interface or a boundary, the electric vectors must satisfy the conditions
\begin{equation} \label{electric boundary conditions}
\mathbf{n} \times \llbracket \mathbbm{e} \rrbracket = \mathbf{0}, \quad \mathbf{n} \cdot \llbracket \mathbbm{d} \rrbracket = \hat{q},
\end{equation}
where $\hat{q}$ is the surface charge density, $\mathbf{n}$ is the unit outward normal to the surface and $\llbracket \bullet \rrbracket$ represents the difference $\left( \bullet^{\mbox{\scriptsize out}} - \bullet^{\mbox{\scriptsize in}} \right)$.

\subsubsection{Linear and angular momentum balance}

The balance of linear momentum in the configuration $\mathcal B_t$ is given in terms of the total Cauchy stress tensor as
\begin{equation}
\mbox{div}\, \boldsymbol{\sigma}^{\text{tot}} + \mathbf{f}_m = \rho \mathbf{a}.
\end{equation}
Here $\boldsymbol{\sigma}^{\text{tot}}$ is the total Cauchy stress tensor that takes both mechanical and electric effects into account, $\mathbf{f}_m$ is the purely mechanical body force, $\rho$ is the mass density, $\mathbf{a}$ is the acceleration, and the divergence operator is taken to operate on the first index of a second order tensor. We refer to Appendix A for a detailed derivation of the balance equations in the context of electroelasticity.

The above equation can be written in referential form using the total Piola--Kirchhoff stress $\mathbf{S}^{\text{tot}} = J \mathbf{F}^{-1} \boldsymbol{\sigma}^{\text{tot}} \mathbf{F}^{-t}$ as
\begin{equation}
\mbox{Div}\left( \mathbf{S}^{\text{tot}} \mathbf{F}^t \right) + \mathbf{f}_M = \rho_r \mathbf{a},
\end{equation}
with $\rho_r = J \rho$ being the referential mass density and $\mathbf{f}_M = J\mathbf{f}_m$ being the referential body force. Note that the tensor $\mathbf{S} = J \mathbf{F}^{-1} \boldsymbol{\sigma}\mathbf{F}^{-t}$ is sometimes also referred to as the `second' Piola--Kirchhoff stress.

The principle of balance of angular momentum renders the Cauchy and the Piola--Kirchhoff stress tensors symmetric
\begin{equation}
 \left( \boldsymbol{\sigma}^{\text{tot}} \right)^t =\boldsymbol{\sigma}^{\text{tot}}, \quad \left(\mathbf{S}^{\text{tot}} \right)^t  = \mathbf{S}^{\text{tot}}.
\end{equation}

The corresponding boundary conditions are given by the equations \eqref{bc cond 54} and \eqref{bc 51}.

\subsubsection{Internal dissipation}

Very often, the EAPs are synthesized from a rubber like polymer.
The polymeric rubber matrix is viscoelastic in nature which leads to energy dissipation on a mechanical deformation. In addition to this, energy dissipation can also occur due to a time-dependent polarisation of the material on application of an electric field.  We consider the possibility that on a sudden application of an electric field (or a potential difference), the electric displacement $\mathbb{D}$, the polarisation $\mathbb P$, and the resulting electric contribution to stress generated in the material evolve with time to reach an equilibrium value. Thus, the two effects need to be modelled appropriately.

To take into account mechanical viscous effects, we assume the existence of an intermediate configuration $\mathcal B_i$ that is, in general, incompatible.
The tangent spaces of $\mathcal B_0$ and $\mathcal B_i$ are related by a second order tensor $\mathbf{F}_v$ that quantifies viscous motion while the tangent spaces of $\mathcal B_i$ and $\mathcal B_t$ are related by a second order tensor $\mathbf{F}_e$ that quantifies elastic distortion.
The configuration $\mathcal B_i$ is in parallel to  the energy-conserving electroelastic deformation from $\mathcal B_0$ to $\mathcal B_t$. This motivates the decomposition of the deformation gradient into an elastic and a viscous part (cf. Lubliner \cite{Lubliner1985} and Reese and Govindjee \cite{Reese1998}) as
\begin{equation} \label{F Fe Fv decomposition}
\mathbf{F} = \mathbf{F}_e \mathbf{F}_v.
\end{equation}
For future use we define the right  and the left Cauchy--Green strain tensors as $\mathbf{C} = \mathbf{F}^t \mathbf{F}$ and  $\mathbf{b} = \mathbf{F} \mathbf{F}^t$, respectively. Corresponding quantities are defined for the viscous parts as $\mathbf{C}_v = \mathbf{F}^t_v \mathbf{F}_v$ and $\mathbf{b}_v = \mathbf{F}_v \mathbf{F}_v^t$.

In order to model the electric dissipation effects, we consider an additive decomposition of the electric field vector  into an `elastic' and a `viscous' part
\begin{equation}
\mathbbm{e} = \mathbbm{e}_e + \mathbbm{e}_v, \quad \mathbb{E}  = \mathbb{E}_e + \mathbb{E}_v,
\end{equation}
which defines a dissipation mechanism in parallel to an energy-conserving mechanism that gives the instantaneous response.
The above additive decomposition of the electric field is motivated by a similar decoupling of the deformation gradient into elastic and viscous parts in the viscoelasticity theory. A similar approach has been taken by Saxena et al.~\cite{Saxena2013a} to model magneto-viscoelastic effects wherein an additive decomposition of the magnetic induction as $\mathbb{B} =\mathbb{B}_e + \mathbb{B}_v$  has been performed.
The behaviour of the internal variables defined above is assumed such that if a constant electric field is applied at time $t=0$, then at that instant $\mathbb{E}_v = \mathbf{0}$ and $\mathbb{E}_e = \mathbb{E}$. As time progresses, $\mathbb{E}_v \rightarrow \mathbb{E}$ and $\mathbb{E}_e \rightarrow \mathbf{0}$.

The material response is assumed to be given by
 an energy density function $\Omega$ that depends on the parameters $\mathbf{C}, \mathbf{C}_v, \mathbb{E}, \mathbb{E}_v$ and the temperature $\vartheta$, (see Appendix A for details).
Taking partial derivatives of $\Omega$ with respect to its arguments and substituting in the thermodynamical  inequality \eqref{SLT last appendix}, we get
\begin{align}
-\left[ \frac{\partial \Omega}{\partial \mathbb E} + \mathbb{D}  \right] \cdot \bdot{\mathbb{E}} - \left[ \frac{\partial \Omega}{\partial \theta} + \rho_r s \right] \bdot{\vartheta} - \left[ \frac{\partial \Omega}{\partial \mathbf{C}} - \frac{1}{2} \mathbf{S} + p \mathbf{C}^{-1} \right] : \bdot{\mathbf{C}} \nonumber \\
- \frac{\partial \Omega}{\partial \mathbf{C}_v} : \bdot{ \mathbf{C}}_v - \frac{\partial \Omega}{\partial \mathbb{E}_v} \cdot \bdot{ \mathbb{E}}_v - \frac{1}{\vartheta} \mathbf{Q} \cdot \mbox{Grad}\, \vartheta \ge 0,
\end{align}
where $p$ is a Lagrange multiplier associated with the constraint of incompressibility and for the sake of brevity we now refer to the total Piola--Kirchhoff stress $\mathbf{S}^{\text{tot}}$ as simply $\mathbf{S}$. A superposed dot represents a material time derivative.

On an application of the procedure due to Coleman and Noll \cite{Coleman1963}, we arrive at the following constitutive relations for the total Piola--Kirchhoff stress, the electric displacement and the specific entropy
\begin{equation} \label{constitutive relations}
\mathbf{S} = 2 \frac{\partial \Omega}{\partial \mathbf{C}} - p \mathbf{C}^{-1}, \quad \mathbb{D} = - \frac{\partial \Omega}{\partial \mathbb E}, \quad s = - \frac{1}{\rho_r} \frac{\partial \Omega}{\partial \vartheta},
\end{equation}
and the dissipation conditions
\begin{equation}
\frac{\partial \Omega}{\partial \mathbf{C}_v} : \bdot{ \mathbf{C}}_v + \frac{\partial \Omega}{\partial \mathbb{E}_v} \cdot \bdot{ \mathbb{E}}_v + \frac{1}{\vartheta} \mathbf{Q} \cdot \mbox{Grad}\, \vartheta \le 0.
\end{equation}

It is noted here that if the incompressibility constraint $(J=1)$ is not imposed, then the constitutive equation for stress is simply
\begin{equation}
 \mathbf{S} = 2 \frac{\partial \Omega}{\partial \mathbf{C}}.
\end{equation}

It is further assumed that the `viscous' electric field, viscous strain and the temperature are independent of each other, thereby reducing the above inequality to the following separate conditions
\begin{equation} \label{sep diss conditions}
\frac{\partial \Omega}{\partial \mathbf{C}_v} : \bdot{ \mathbf{C}}_v \le 0, \quad \frac{\partial \Omega}{\partial \mathbb{E}_v} \cdot \bdot{\mathbb{E}}_v \le 0, \quad  \mathbf{Q} \cdot \mbox{Grad}\, \vartheta \le 0.
\end{equation}
These conditions must be satisfied by all processes to be thermodynamically admissible. Isothermal conditions are assumed henceforth in this paper and we remove the dependence on the temperature $\vartheta$.

A common technique useful for performing numerical computation  (\cite{Reese1998}, \cite{Holzapfel1996a}) is to decompose the total energy into an equilibrium part associated with the direct deformation from $\mathcal B_0$ to $\mathcal B_t$, and a non-equilibrium part due to the internal variable $\mathbb E_v$ and the elastic deformation from $\mathcal B_i$ to $\mathcal B_t$.
\begin{equation}
\Omega \left( \mathbf{C}, \mathbf{C}_v, \mathbb{E}, \mathbb E_v \right) = \Omega_e \left( \mathbf{C}, \mathbb{E} \right) + \Omega_v \left( \mathbf{C}, \mathbf{C}_v, \mathbb E, \mathbb E_v \right).
\end{equation}

The above simplification is used to write the dissipation inequalities \eqref{sep diss conditions}$_{1,2}$ as
\begin{equation} \label{thermo constraints final}
\frac{\partial \Omega_v}{\partial \mathbf{C}_v} : \bdot{ \mathbf{C}}_v \le 0, \quad \frac{\partial \Omega_v}{\partial \mathbb{E}_v} \cdot \bdot{ \mathbb{E}}_v \le 0.
\end{equation}

\subsection{Specialised constitutive laws}

To obtain numerical results in order to understand the physical implications of the several aspects of the developed theory, we use a prototype energy function that is a generalisation of the isotropic neo-Hookean function to electroelasticity and is given by
\begin{equation} \label{energy omega E}
\Omega_e = \frac{\mu_e}{2}  \left[ I_1-3 \right] + m_e I_4 + n_e I_5,
\end{equation}
where the scalar invariants $I_1, I_4$ and $I_5$ are defined as
\begin{equation}
I_1  = \mathbf{C}: \mathbf{I}, \quad I_4 = \left[ \mathbb E \otimes \mathbb{E} \right]: \mathbf{I} , \quad I_5 = \big[ \mathbb{E} \otimes \left[ \mathbf{C}^{-1} \mathbb{E} \right] \big]: \mathbf{I}.
\end{equation}
The parameter $\mu_e$ is the shear modulus in the absence of electric fields, $m_e$ and $ n_e$  are electroelastic coupling parameters with $m_e/\varepsilon_0$ and $ n_e/\varepsilon_0 $ being dimensionless, and $\mathbf{I}$ is the identity tensor in $\mathcal B_0$.

For the non-equilibrium part of the stored energy, a similar (but with one different electro-mechanical coupling term) form as that of $\Omega_e$ in \eqref{energy omega E} is chosen in a way that the energy depends only on the non-equilibrium `elastic' variables $\mathbf{C}_e = \mathbf{F}_v^{-t} \mathbf{C} \mathbf{F}_v^{-1}$ and $\mathbb{E}_e = \mathbb{E} - \mathbb{E}_v$. 
The reason and the procedure of such a simplification is discussed in \cite{Saxena2013a}.
\begin{align}
&\Omega_v = \frac{\mu_v}{2}  \left[ \mathbf{C} : \mathbf{C}_v^{-1} - 3 \right]  + m_v \big[ \left[ \mathbb{E} - \mathbb{E}_v \right] \otimes \left[ \mathbb{E} - \mathbb{E}_v  \right] \big] : \mathbf{I} \nonumber \\
& + n_v \Big[ \big[ \mathbf{C} \left[ \mathbb{E} - \mathbb{E}_v \right] \big] \otimes \big[ \mathbf{C} \left[ \mathbb{E} - \mathbb{E}_v  \right] \big] \Big] : \mathbf{I}. \label{energy omega V}
\end{align}
Here $\mu_v, m_v$ and $n_v$ are the non-equilibrium parameters with dimensions similar to their electroelastic counterparts defined in \eqref{energy omega E}.

The evolution equation for the `viscous' electric field vector is taken to be
\begin{align}
\bdot{\mathbb E}_v &= - \frac{1}{\xi T_e} \frac{\partial \Omega_v}{\partial \mathbb{E}_v} ,\\
& = \frac{2 }{\xi  T_e} \left[m_v \mathbf{I} + n_v \mathbf{C}^2 \right] \left[ \mathbb{E} - \mathbb{E}_v \right], \label{elec evolve}
\end{align}
where we have taken $T_e$ as a time constant to measure evolution of `viscous' electric field and $\xi$ is a scaling parameter.
For the viscous deformation, we use the evolution equation given by Koprowski-Theiss et al.~\cite{Koprowski-Theiss2011} that is a simplification of a general form used by Lion \cite{Lion1997}
\begin{equation}
\bdot{\mathbf{C}}_v = \frac{1}{T_v} \left[ \mathbf{C} - \frac{1}{3} \left[ \mathbf{C} : \mathbf{C}_v^{-1} \right] \mathbf{C}_v \right]. \label{deform evolve}
\end{equation}
Both the evolution equations \eqref{elec evolve} and \eqref{deform evolve} satisfy the thermodynamic constraints \eqref{thermo constraints final}$_{1,2}$. They are also defined such that the viscous quantities $\mathbb{E}_v$ and $\mathbf{C}_v$ stop evolving as the system reaches equilibrium.

For the above defined energy density functions, the total Cauchy stress $\boldsymbol{\sigma} = J^{-1} \mathbf{F} \mathbf{S} \mathbf{F}^t$ is given as
\begin{equation} \label{sigma = se+sv+pi}
\boldsymbol{\sigma} = \boldsymbol{\sigma}^e + \boldsymbol{\sigma}^v + p \boldsymbol{i},
\end{equation}
where $\boldsymbol{i}$ is the identity tensor in $\mathcal B_t$ and
\begin{equation} \label{taue value}
\boldsymbol{\sigma}^e = \mu_e   \mathbf{b} - 2 n_e \mathbbm{e} \otimes \mathbbm{e} ,
\end{equation}
\begin{equation} \label{tauv value}
 \boldsymbol{\sigma}^v = \mu_v   \mathbf{b}_e + 2 n_v \left[ \left[ \mathbf{b}\, \mathbbm{e}_e \right] \otimes \left[ \mathbf{b}^2 \mathbbm{e}_e \right] + \left[ \mathbf{b}^2 \mathbbm{e}_e \right] \otimes \left[ \mathbf{b}\, \mathbbm{e}_e \right] \right].
\end{equation}

The total electric displacement $\mathbbm{d}$ is given as
\begin{equation}
\mathbbm{d} = \mathbbm{d}^e + \mathbbm{d}^v,
\end{equation}
\begin{equation}
\mathbbm{d}^e = -2 m_e \mathbf{b}\, \mathbbm{e} - 2 n_e \mathbbm{e},
\end{equation}
\begin{equation} \label{dv final expr}
\mathbbm{d}^v = -2 m_v \mathbf{b}\, \mathbbm{e}_e - 2 n_v \mathbf{b}^3 \mathbbm{e}_e.
\end{equation}

We now use the above defined energies and specialised constitutive laws to perform numerial calculations in the following sections.

\section{Numerical examples}
In order to understand the physical behaviour predicted by the model developed in the sections above, we consider some numerical examples. By prescribing the deformation and the electric field at a point, we calculate the evolution of stress and electric displacement in the material. The dependence of the total Cauchy stress and the electric displacement on modelling parameters, applied stretch, and electric potential difference or electric field is demonstrated graphically.

\subsection{No deformation}

In the first case, in order to highlight the effects of non-equilibrium electric field, we consider no deformation but existence of an electric field. Let the test specimen be held fixed ($\mathbf{C} = \mathbf{I}$) using appropriate boundary tractions and a sudden but constant electric field is applied at time $t=0$. This leads to the generation of a viscous overstress and an electric displacement, both of which settle down to equilibrium values with time. 
The following values of the material parameters  are used for calculations in this case

\begin{align}
 \mu_e = 5 \times 10^6 \; \text{MPa}, \quad \mu_v = 2 \times 10^6 \; \text{MPa}, \quad m_e = -10 \; \text{N/V}^2, \quad \xi = 1  \nonumber \\
 n_e = -6 \; \text{N/V}^2, \quad T_e = 1 \; \mbox{s}, \quad T_v = 50 \; \mbox{s}, \quad m_v = - m_e, \quad n_v =- n_e .
\end{align}

The electric field is given by
\begin{equation}
\mathbb{E}_2 = \mathbb{E}_3 = 0, \quad
  \mathbb{E}_1 = \begin{cases}
    0, & \mbox{for}\;\; t<0,\\
    3 \times 10^2 \, \mbox{V/m}, & \mbox{for}\; \; t\ge 0.
  \end{cases}
\end{equation}

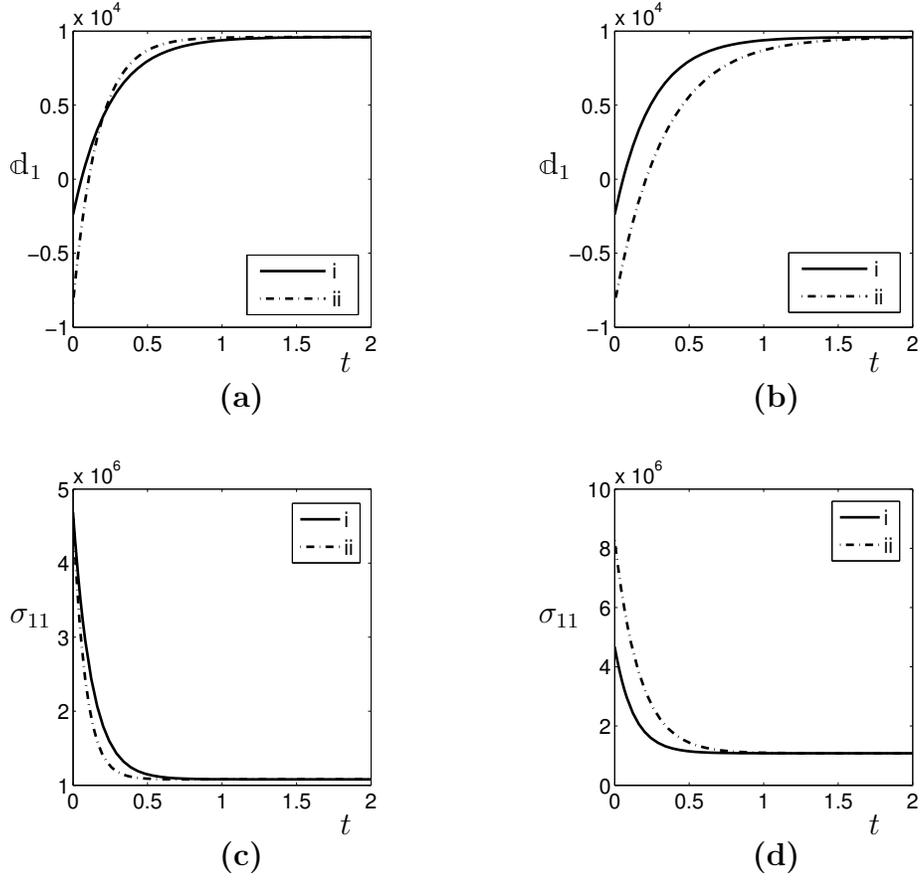
\begin{figure}
\begin{center}
\begin{tabular}{c c}
\includegraphics[scale=0.5]{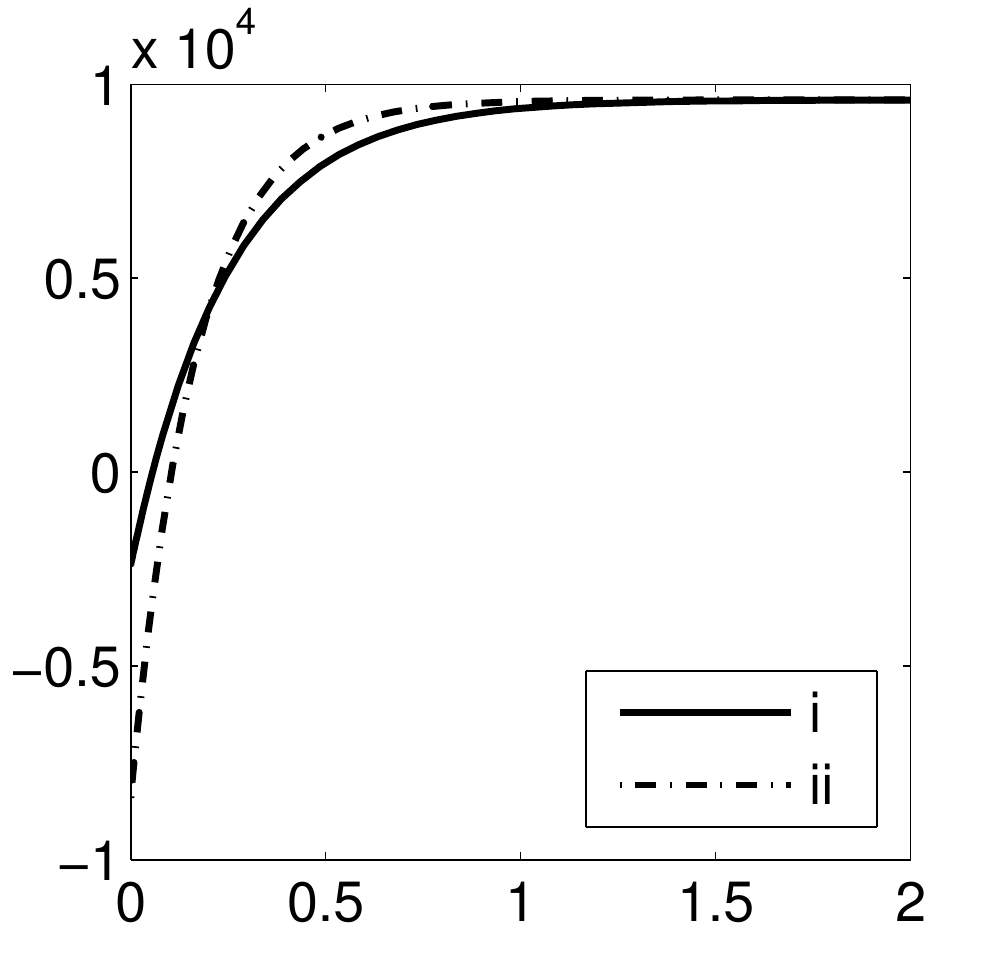} \hspace{20pt} & \hspace{20pt} \includegraphics[scale=0.5]{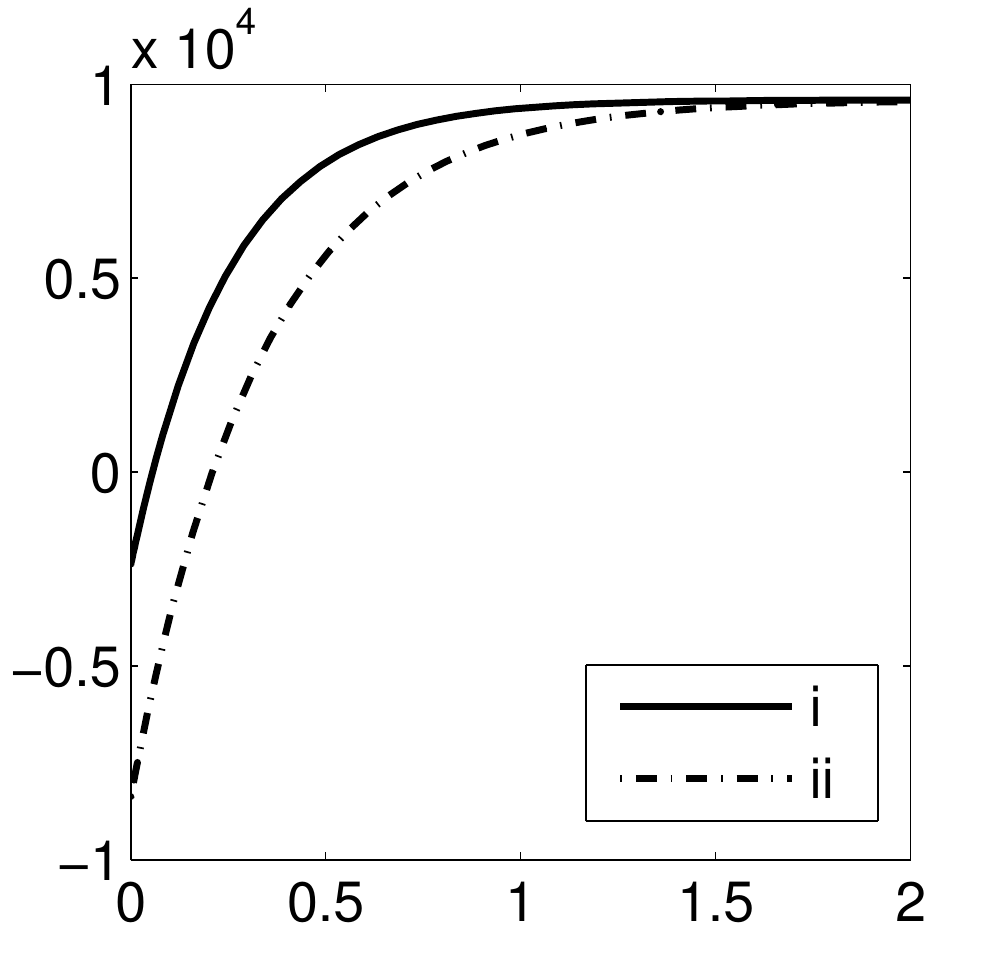} \\
\textbf{(a)} \hspace{35pt}& \hspace{35pt} \textbf{(b)}\\
\\
\includegraphics[scale=0.5]{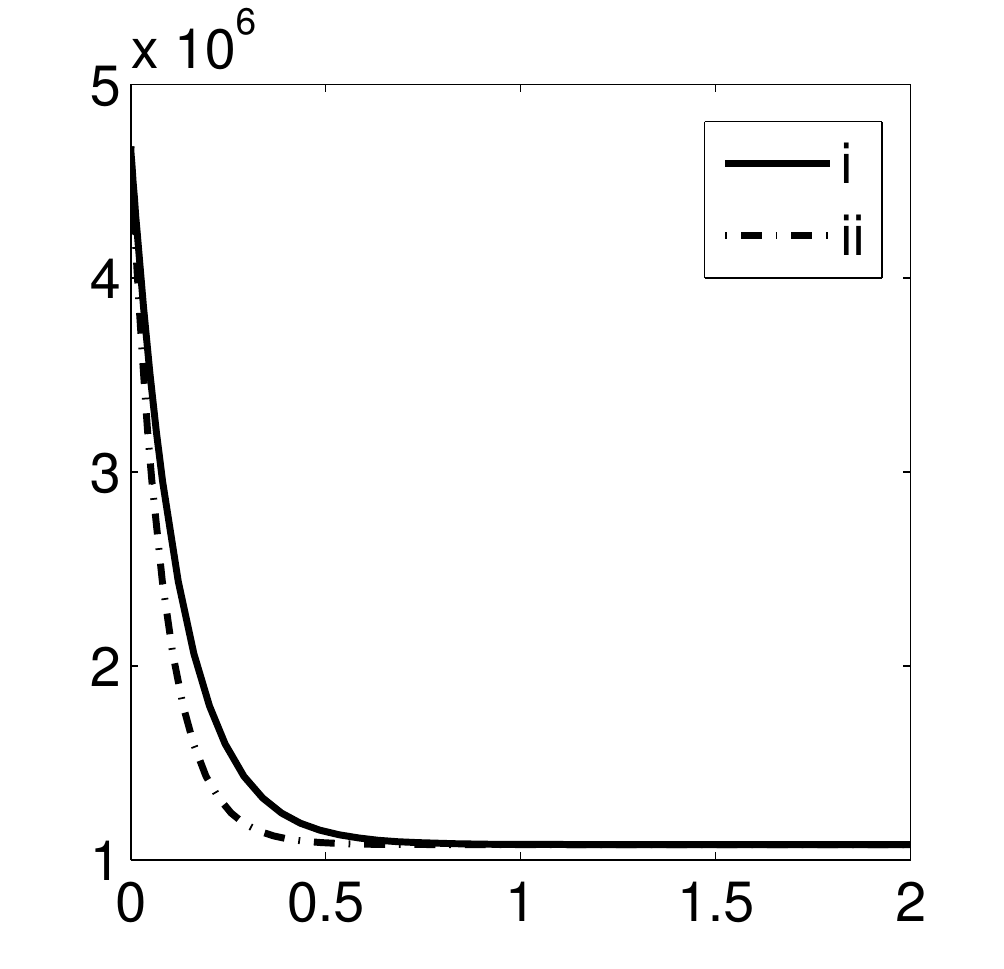} \hspace{20pt} & \hspace{20pt} \includegraphics[scale=0.5]{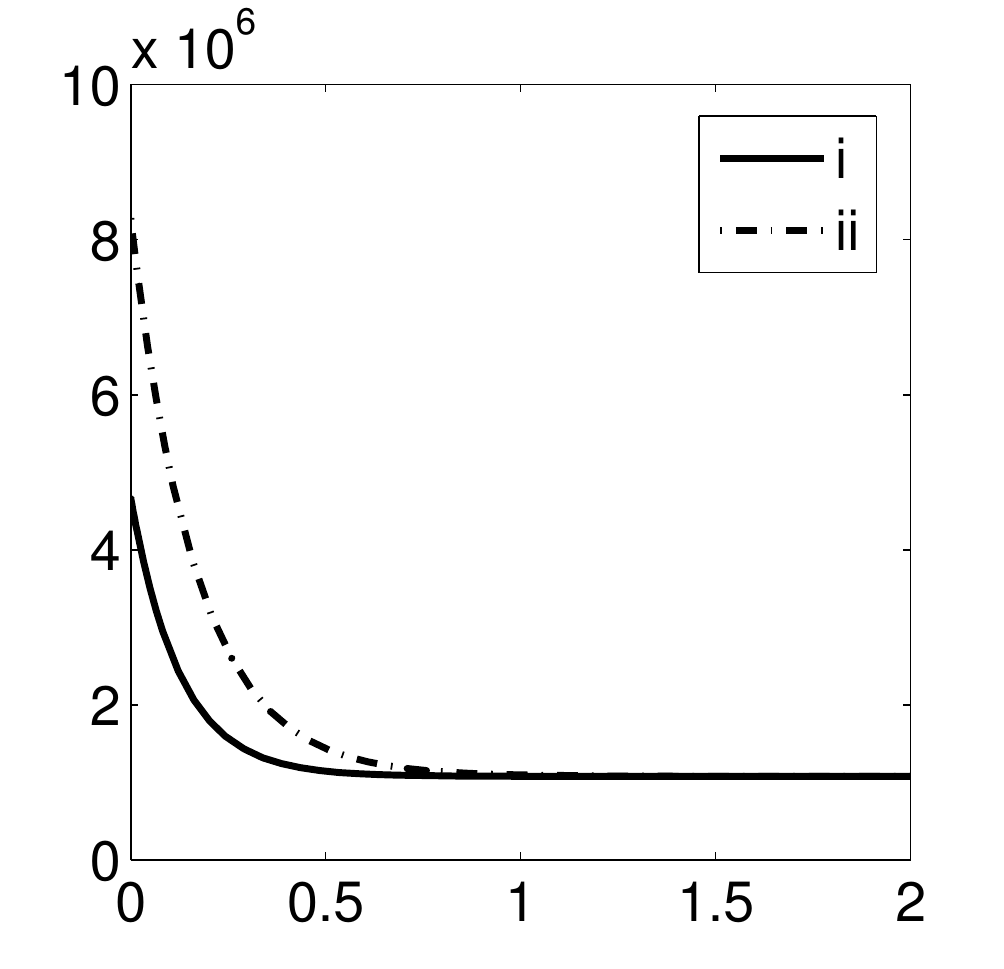}\\
\textbf{(c)} \hspace{35pt}& \hspace{35pt} \textbf{(d)}
\end{tabular}
\end{center}
\begin{picture}(0,0)(0,0)
\put(140,30){$t$} \put(140,205){$t$} \put(340, 30){$t$} \put(340, 205){$t$}
\put(15, 280){$\mathbbm{d}_1$} \put(215, 280){$\mathbbm{d}_1$}
\put(15, 110){$\sigma_{11}$} \put(215, 110){$\sigma_{11}$}
\end{picture}
\caption{Evolution of (\textbf{a,b}) electric displacement $\mathbbm{d}_1$ (Nm$^{-1}$V$^{-1}$) and (\textbf{c,d})~the total Cauchy stress $\sigma_{11}$ (N/m$^2$) with time for two different values of $m_v$ and $n_v$: \textbf{(a,c)} (i) $m_v = 3$~N/V$^2$, (ii) $m_v = 10$~N/V$^2$; \textbf{(b,d)} (i)~$n_v = 3$~N/V$^2$, (ii)~$n_v = 10$~N/V$^2$.}
\label{fig: no def 1}
\end{figure}

Time integrations of the evolution equations are performed using a modified Runge-Kutta scheme implemented in the \texttt{ode45} solver of MATLAB. It is observed that an `electro-viscous' stress is developed in the material at time $t=0$ when the electric field is switched on. The stress decays with time due to evolution of the `viscous' electric field $\mathbb{E}_v$ and reaches an equilibrium value. The rate of decay and  the initial non-equilibrium value depend on the material parameters $m_v$ and $n_v$. As observed from Fig.~\ref{fig: no def 1}(c), changing the value of $m_v$ does not change the initial value of stress $\sigma_{11}$ but a higher $m_v$ causes a faster decay of stress to its equilibrium value.
It's seen from \ref{fig: no def 1}(d) that a higher value of $n_v$ results in a higher value of initial stress while slowing the decay process.

Similar to the non-equilibrium stress, a `viscous' electric displacement is developed in the material at $t=0$ which evolves with time to achieve an equilibrium value. As is seen from Figs.~\ref{fig: no def 1}(a) and \ref{fig: no def 1}(b), a higher value of either $m_v$ or $n_v$ leads to a smaller initial value of electric displacement. A higher $m_v$ leads to a faster evolution while the reverse happens for a higher $n_v$.

\subsection{Uniaxial deformation}

In this case, we apply both the uniaxial tension and the electric field at $t=0$ as a step function. The stretch $\lambda_1$ and the electric field $\mathbb E_1$ are both applied in the same direction.

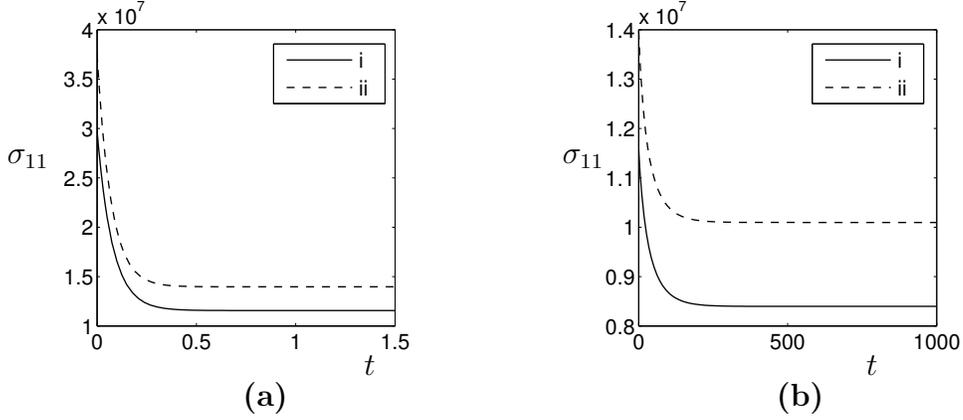
\begin{figure}
\begin{center}
\begin{tabular}{c c}
\includegraphics[scale=0.5]{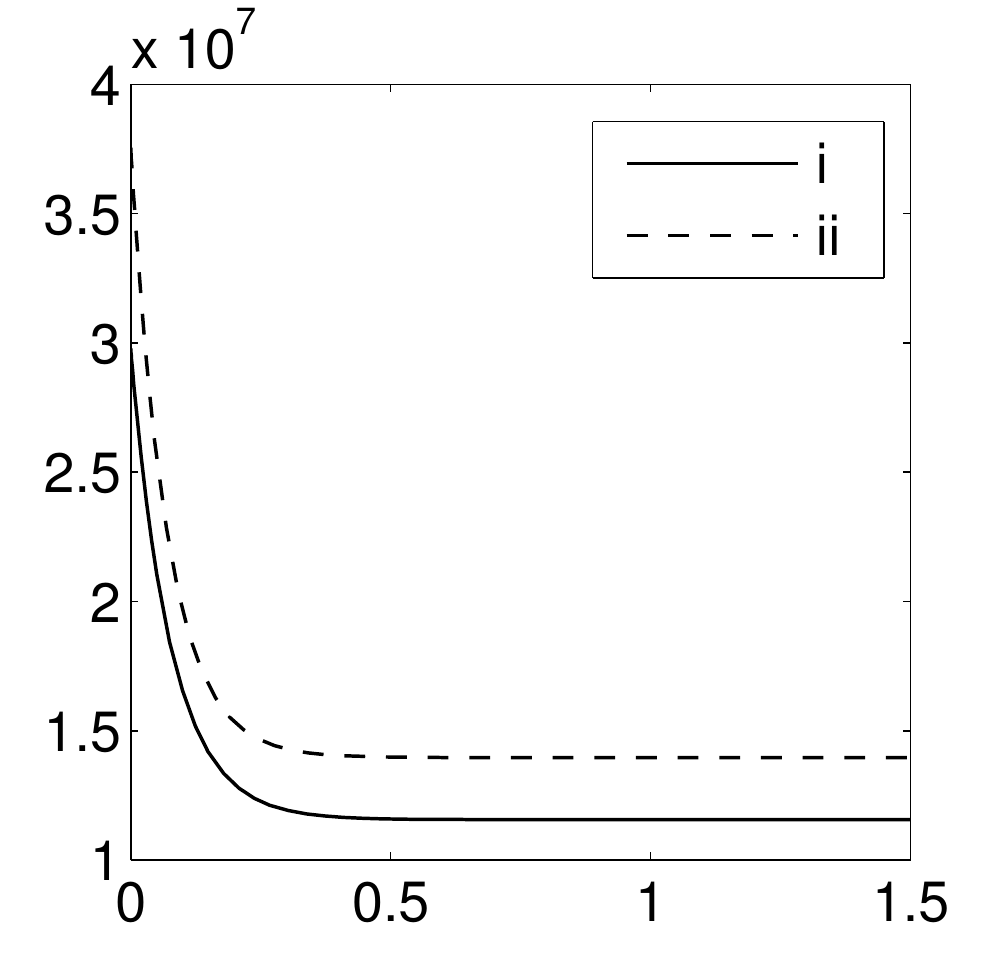} \hspace{20pt} & \hspace{20pt} \includegraphics[scale=0.5]{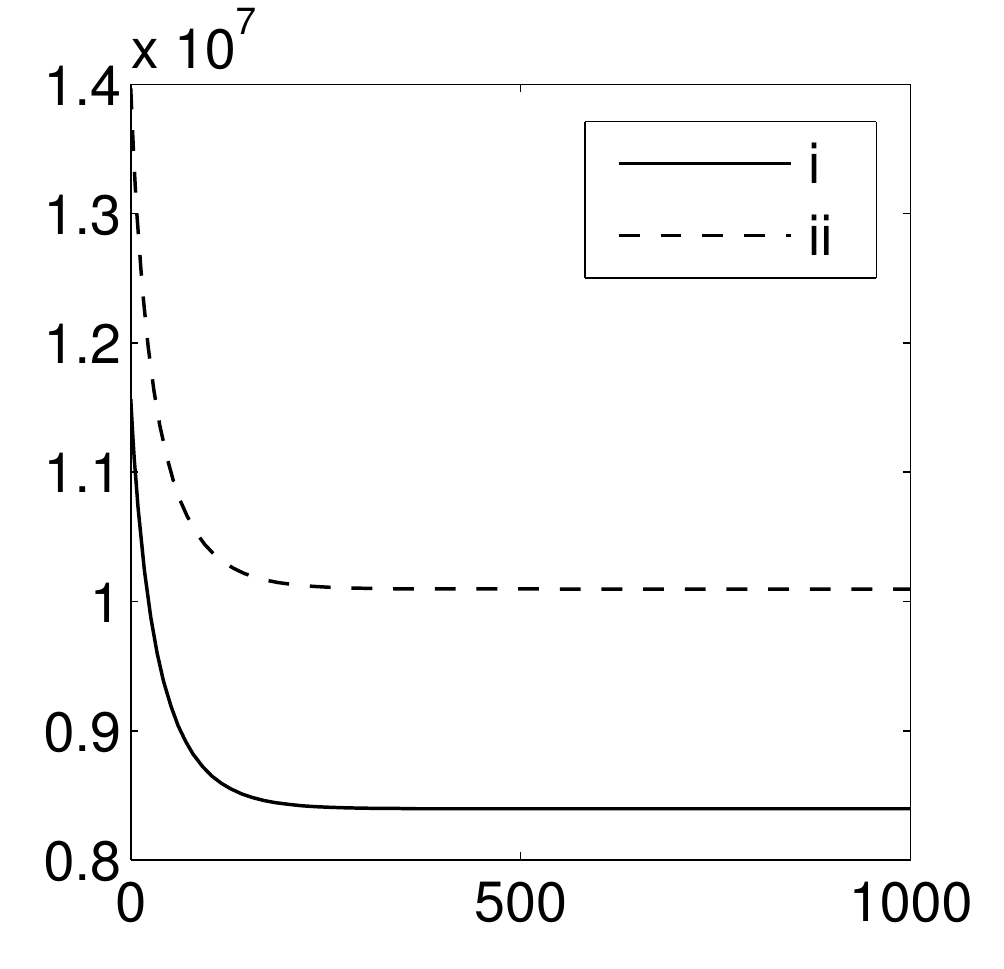} \\
\textbf{(a)} \hspace{35pt}& \hspace{35pt} \textbf{(b)}\\
\end{tabular}
\end{center}
\begin{picture}(0,0)(0,0)
\put(140,30){$t$} \put(330,30){$t$}
\put(5, 110){$\sigma_{11}$} \put(215, 110){$\sigma_{11}$}
\end{picture}
\caption{Uniaxial loading: Evolution of the total Cauchy stress $\sigma_{11}$ (N/m$^2$) with time $t$ (s) for \textbf{(a)} short time-scale and \textbf{(b)} large time-scale. (i) $\lambda_1 = 1.5$, (ii) $\lambda_1 = 1.6$.}
\label{fig: uniaxial}
\end{figure}

The viscous overstress decays at two time scales --  the mechanical part corresponding to evolution of $\mathbf{C}_v$ and  the electric part corresponding to evolution of $\mathbb{E}_v$ as shown in Fig.~\ref{fig: uniaxial}(a) and \ref{fig: uniaxial}(b) where the curves are plotted for two different values of the initial deformation $\lambda_1 = 1.5$ and $\lambda_1 = 1.6$, respectively.
As expected, a higher $\lambda_1$ results in a higher value of stress but it also results in a faster evolution of $\sigma_{11}$.
 As shown, the end point of a curve in Fig.~\ref{fig: uniaxial}(a) has the same value as the starting point of the corresponding curve in Fig.~\ref{fig: uniaxial}(b) since they both correspond to the same loading conditions.

\subsection{Time dependent electric field}
In this section, we study the effect of a time-varying electric field on the induced total stress and the overall dielectric displacement in an undeformed material. The specimen is taken to be fixed at zero deformation and an electric field is applied at time $t=0$ in the $x_1$ direction with a constant rate until a value $\mathbbm{e}_1 = 300$ V/m is obtained. The electric field is then reduced with the same rate until it reduces back to zero. Numerical results for this case are shown in Figs.~\ref{fig: edot edot} and \ref{fig: edot 2} and are obtained for the values $\dot{\mathbbm{e}}_1 = \{\pm 200, \pm 400 \} \; \text{V/m-s}$ in Fig.~\ref{fig: edot edot} and $\dot{\mathbbm{e}}_1 =\pm 300 \; \text{V/m-s}$ in Fig.~\ref{fig: edot 2}.

It is observed from Fig.~\ref{fig: edot edot}(a) that the electric displacement increases with an increasing electric field in a nonlinear fashion and returns back through a different path, eventually reaching a non-zero value of $\mathbbm{d}$ for  zero $\mathbbm{e}$. If the electric field is held at zero after this time, the electric displacement would gradually relax to zero showing a behaviour similar to that in Fig.~\ref{fig: no def 1}(a). A higher rate ($\dot{\mathbbm{e}}_1$) of the electric field leads to lower extreme values of $\mathbbm{d}$ but a larger area enclosed within the curve leading to a higher energy loss per cycle. Another key observation is that the maximum value of dielectric displacement $\mathbbm{d}$ is obtained on the unloading curve just at the beginning of the unloading part of the cycle. This is because the evolution of $\mathbb{E}_v$ slightly lags behind the instantaneous changes in $\mathbbm{e}$. The total Cauchy stress in Fig.~\ref{fig: edot edot}(b) varies almost quadratically with $\mathbbm{e}$, slight hysteresis being observed only at higher rates of electric field. It is also observed that the value of $\sigma_{11}$ on the loading curve is usually higher than that on the unloading curve for larger values of $\mathbbm{e}_1$.

Similar curves for different values of the material parameters $n_v$ and $T_e$ are plotted in Figs.~\ref{fig: edot 2}(a,b) and \ref{fig: edot 2}(c,d), respectively.

A higher $n_v$ increases the maximum value obtained by $\mathbbm{d}$ and results in a larger area under the curve thus increasing the energy dissipated per cycle. The Cauchy stress $\sigma_{11}$ remains largely unchanged with the value of $n_v$. Increasing the value of $T_e$ lowers the peak value of the dielectric displacement but increases that of the total stress. A higher $T_e$ increases the amount of hysteresis in each case.


\begin{figure}
\begin{center}
\begin{tabular}{c c}
\includegraphics[scale=0.5]{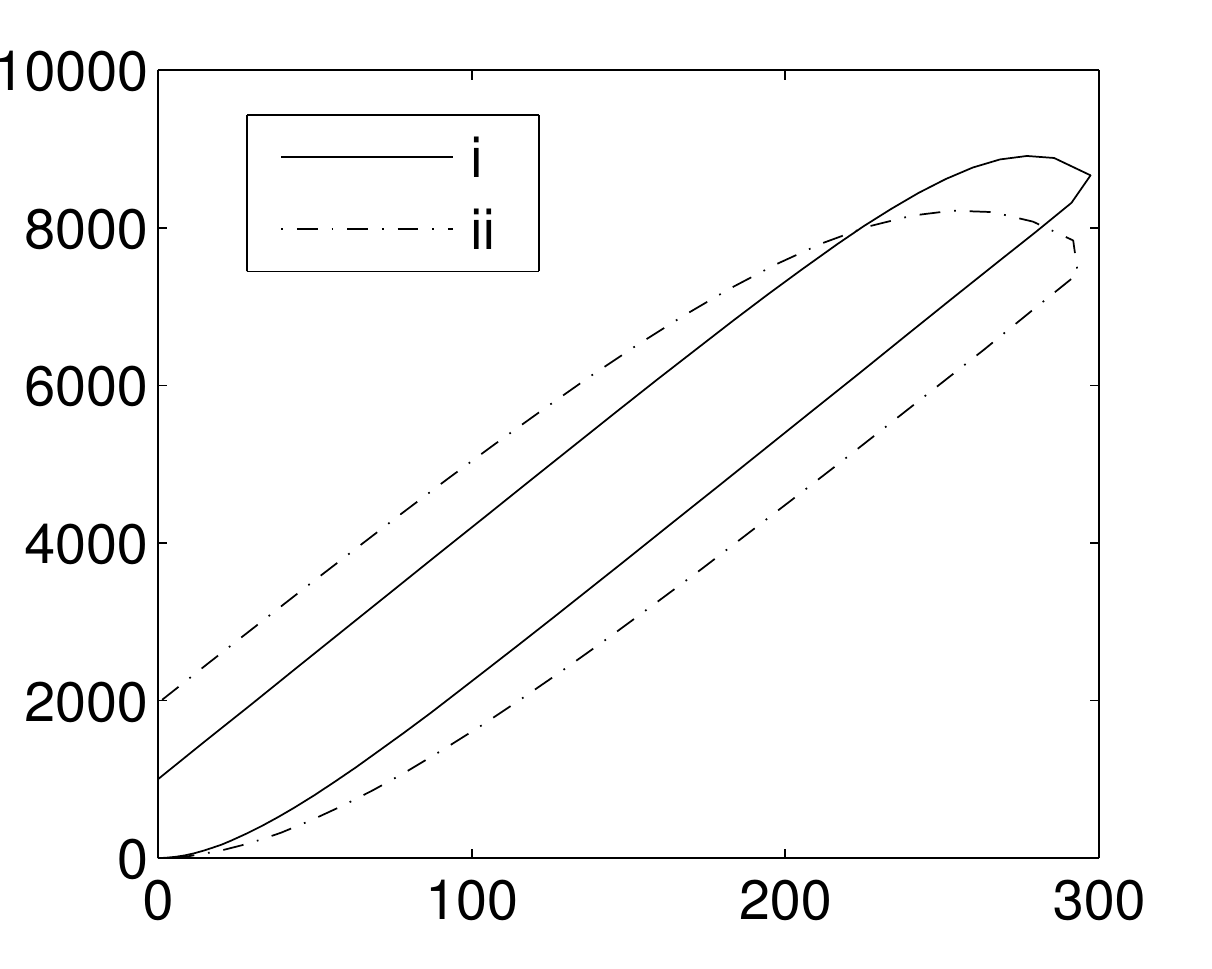} \hspace{10pt} & \hspace{10pt} \includegraphics[scale=0.5]{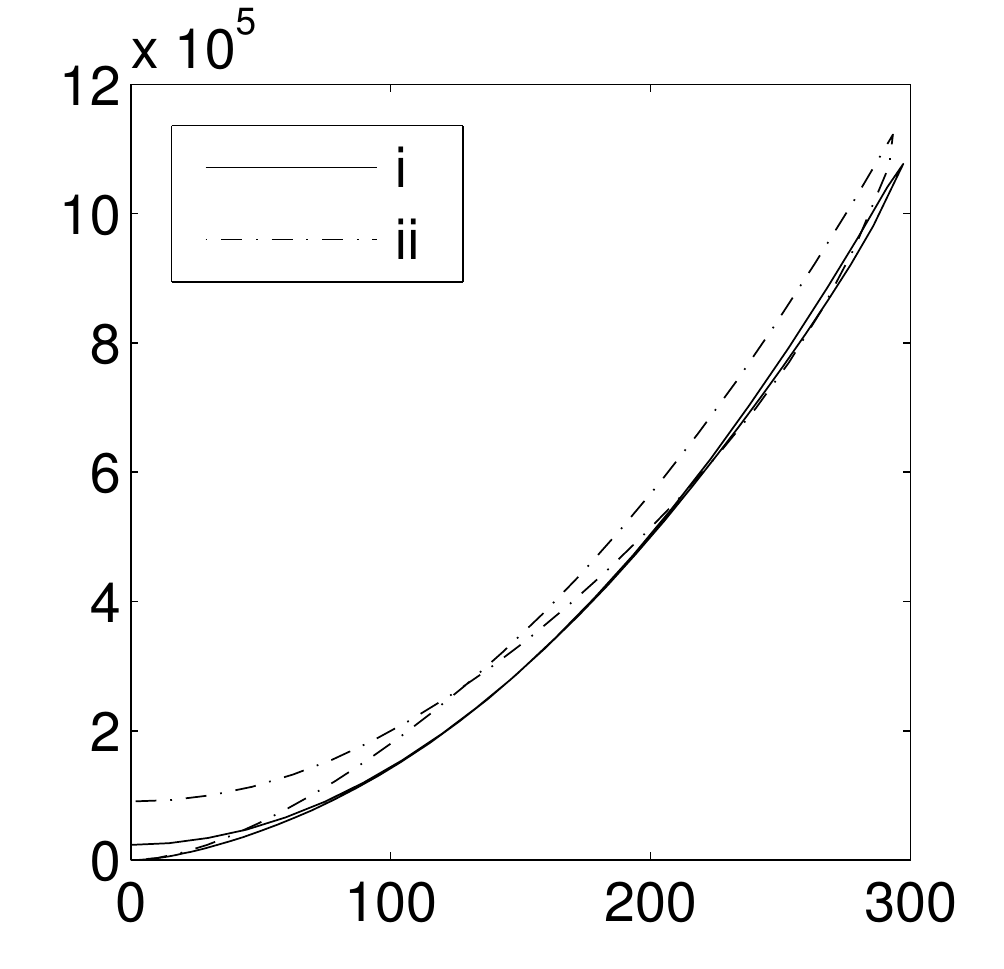} \\
\textbf{(a)} \hspace{35pt}& \hspace{35pt} \textbf{(b)}\\
\end{tabular}
\end{center}
\begin{picture}(0,0)(0,0)
\put(150,30){$\mathbbm{e}_1$} \put(340,30){$\mathbbm{e}_1$}
\put(-5, 110){$\mathbbm{d}_{1}$} \put(220, 110){$\sigma_{11}$}
\end{picture}
\caption{Variation of \textbf{(a)} the electric displacement $\mathbbm{d}_1$ and \textbf{(b)} total Cauchy stress $\sigma_{11}$~(N/m$^2$) with electric field $\mathbbm{e}_1$: (i) $\dot{\mathbbm{e}}_1 = \pm 200$~V/m-s, (ii) $\dot{\mathbbm{e}}_1 = \pm 400$~V/m-s.}
\label{fig: edot edot}
\end{figure}
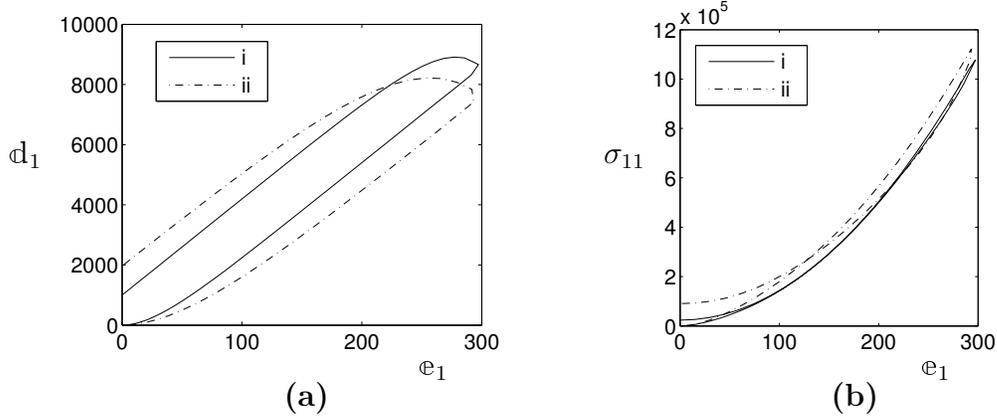
\begin{figure}
\begin{center}
\begin{tabular}{c c}
\includegraphics[scale=0.5]{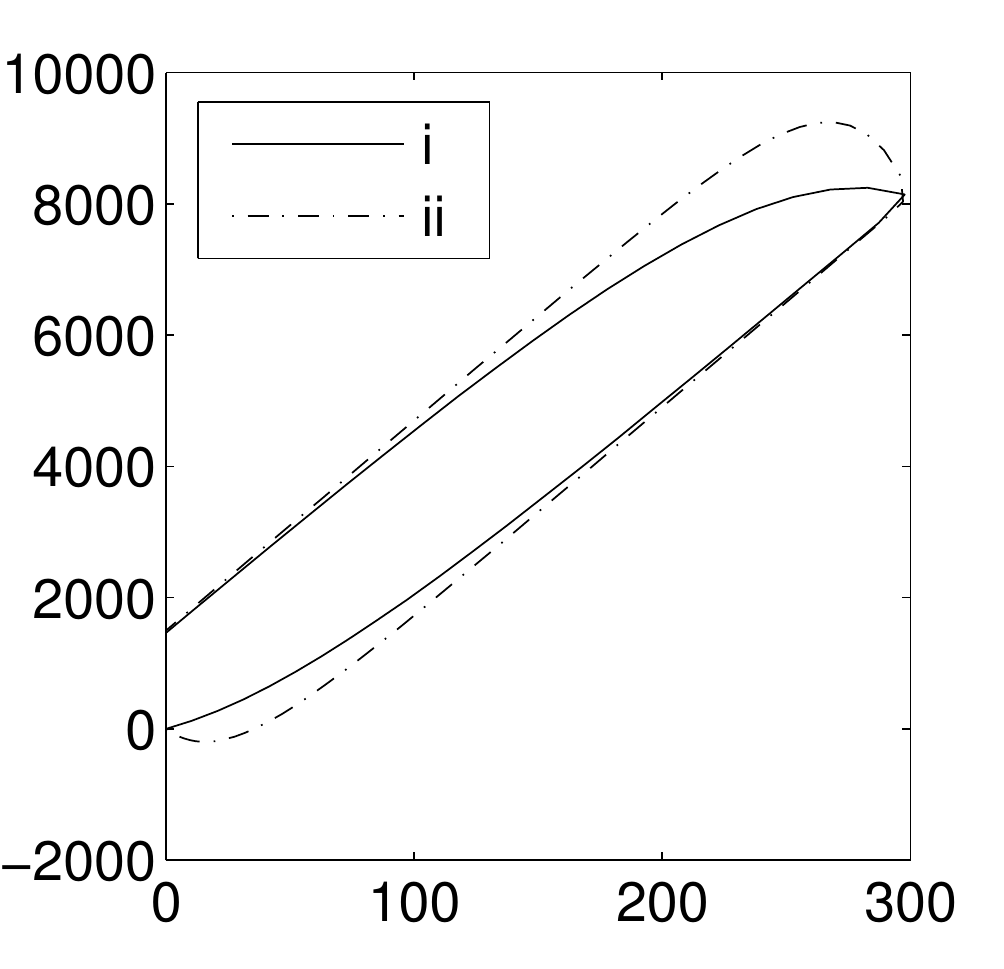} \hspace{20pt} & \hspace{20pt} \includegraphics[scale=0.5]{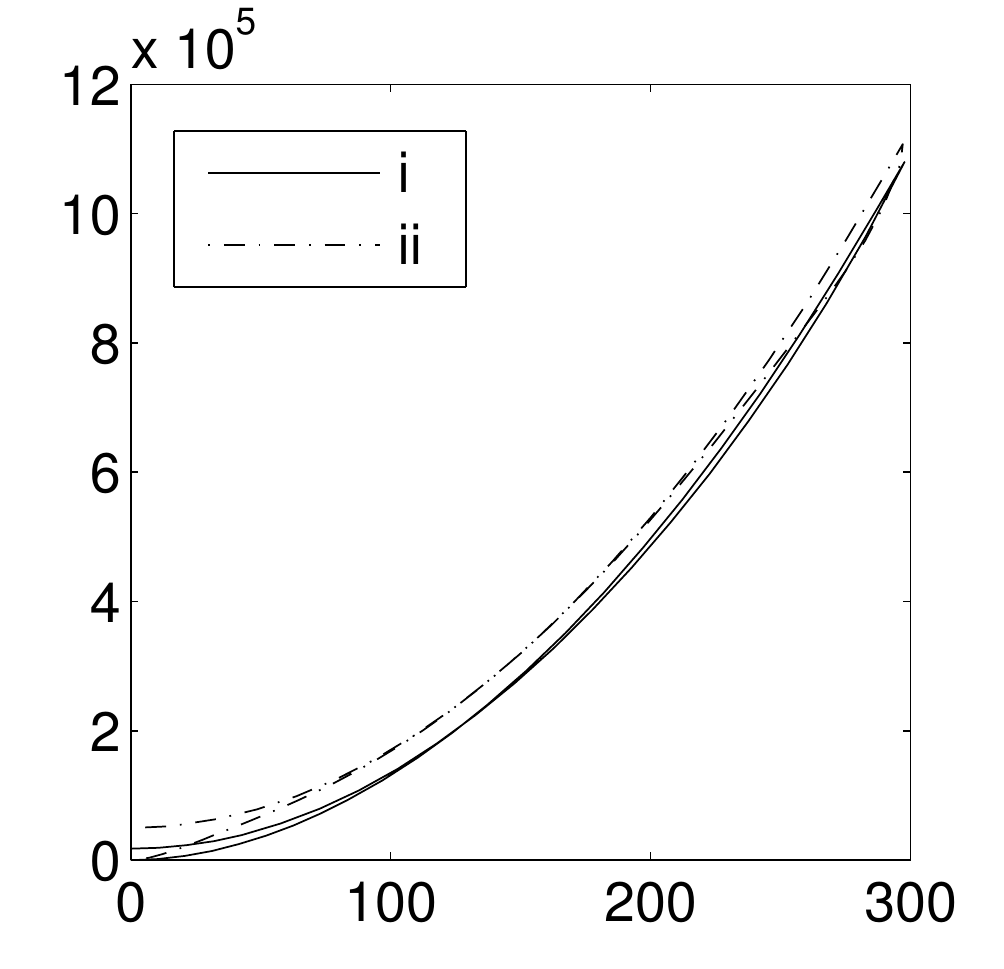} \\
\textbf{(a)} \hspace{35pt}& \hspace{35pt} \textbf{(b)}\\
\\
\includegraphics[scale=0.5]{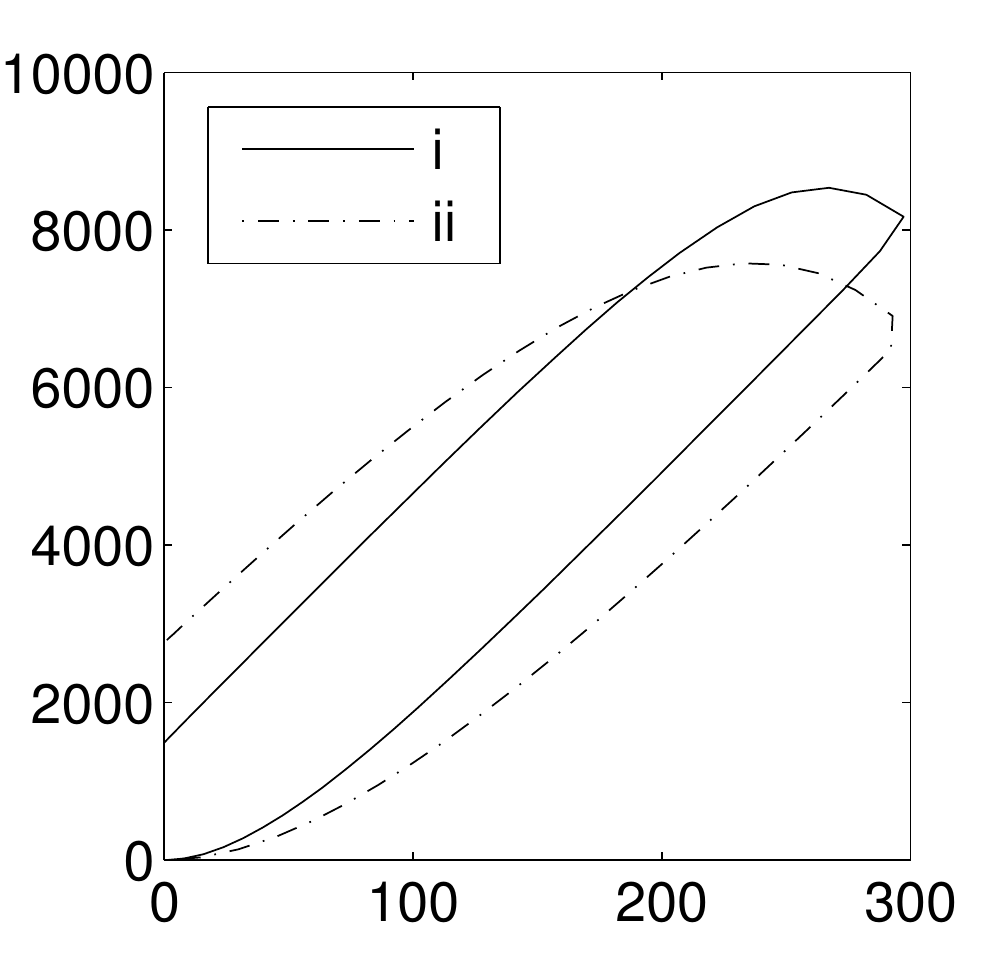} \hspace{20pt} & \hspace{20pt} \includegraphics[scale=0.5]{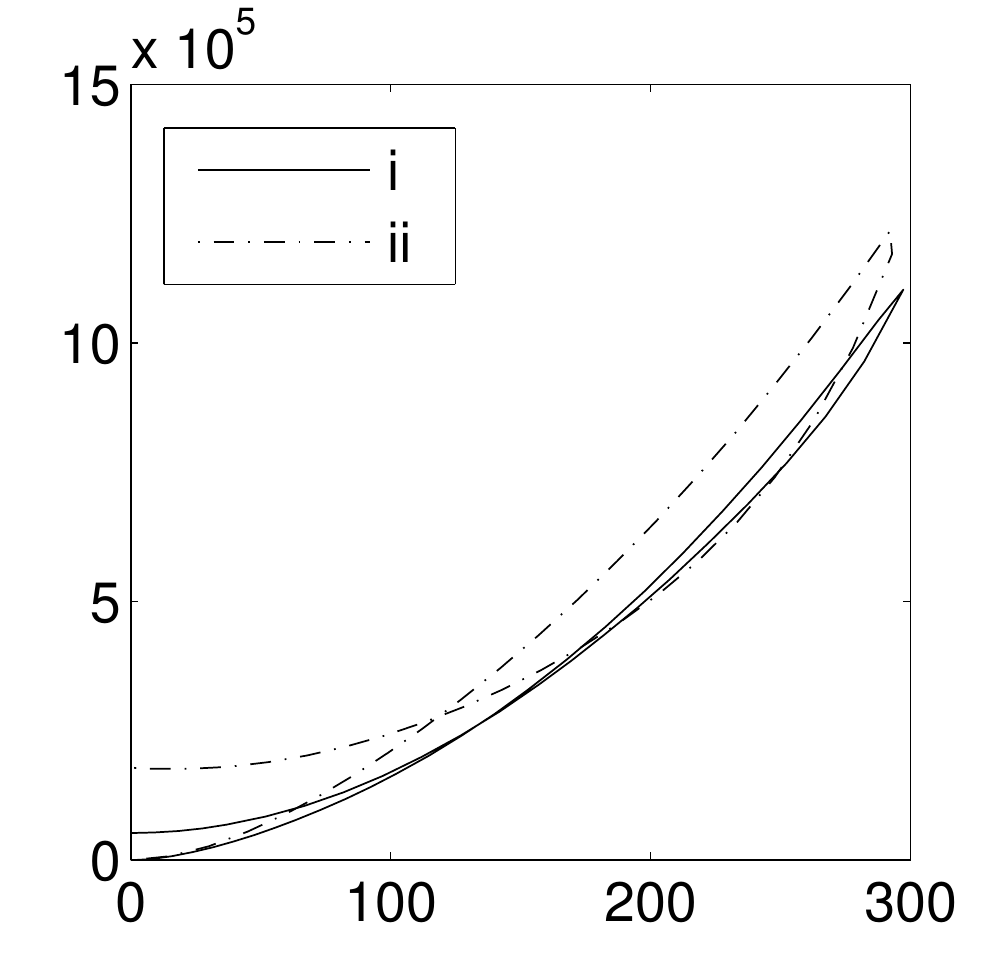}\\
\textbf{(c)} \hspace{35pt}& \hspace{35pt} \textbf{(d)}
\end{tabular}
\end{center}
\begin{picture}(0,0)(0,0)
\put(130,30){$\mathbbm{e}_1$} \put(130,205){$\mathbbm{e}_1$} \put(330, 30){$\mathbbm{e}_1$} \put(330, 205){$\mathbbm{e}_1$}
\put(5, 270){$\mathbbm{d}_1$} \put(215, 290){$\sigma_{11}$}
\put(5, 110){$\mathbbm{d}_{1}$} \put(215, 120){$\sigma_{11}$}
\end{picture}
\caption{Variation of (\textbf{a,c}) the electric displacement $\mathbbm{d}_1$ (Nm$^{-1}$V$^{-1}$) and (\textbf{b,d}) the total Cauchy stress $\sigma_{11}$ (N/m$^2$) with electric field: \textbf{(a.b)} (i) $n_v = 1$~N/V$^2$, (ii) $n_v = 20$~N/V$^2$; \textbf{(c,d)} (i) $T_e = 5$~s, (ii) $T_e = 10$~s.}
\label{fig: edot 2}
\end{figure}
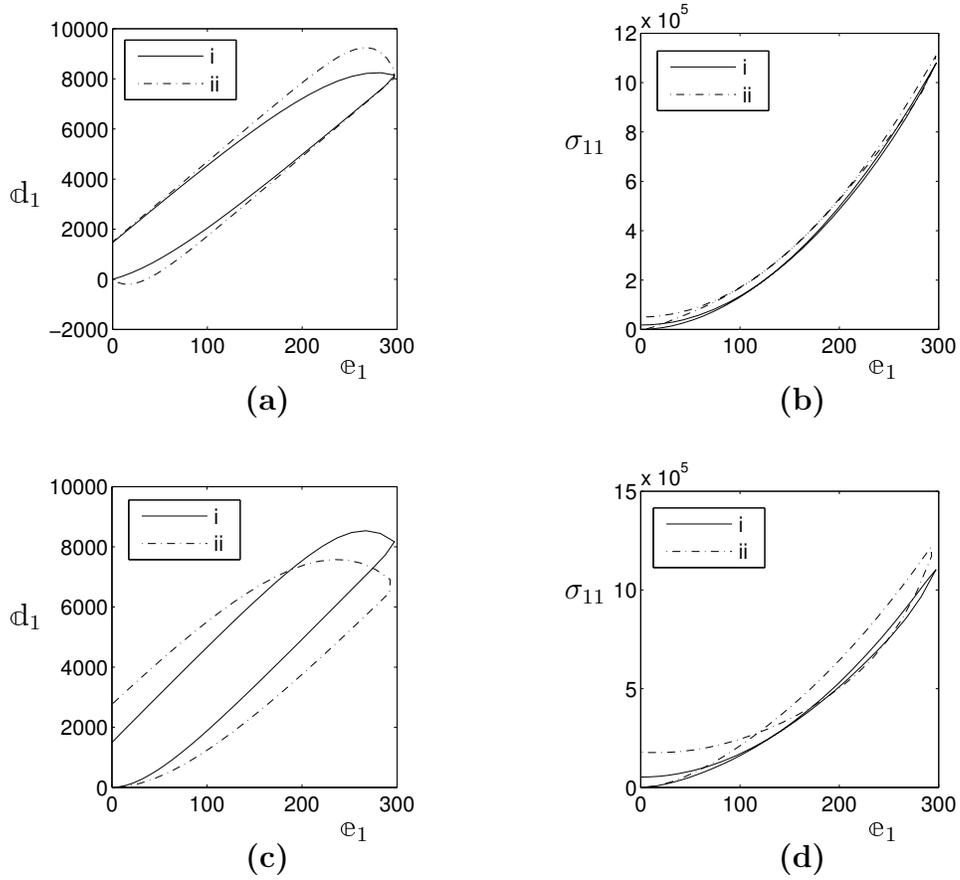

\subsection{Time dependent shear}
In this case, we consider the deformation to be given by
\begin{equation}
 r = R, \quad \theta = \Theta + \alpha Z , \quad z =  Z,
\end{equation}
This is an effective deformation obtained by torsion of a cylinder by constraining the plane cross-sections to remain plane. It is typically obtained experimentally in a disc deformed using a rheometer. Locally, the deformation is a plane strain where $\alpha$ depends on time and is given by $\alpha = \alpha_0 \, \mbox{sin}\, \omega t$. The applied Lagrangian electric field is in the $z$ direction given as $\mathbb{E} = \{0, 0, E_3 \}^t$. Let $(1,2,3)$ correspond to $(R, \Theta, Z)$ in the cylindrical coordinate system, then the deformation gradient $\mathbf{F}$ and the left Cauchy-green strain tensor $\mathbf{b}$ are given as
\begin{equation}
 [\mathbf{F}] = \left[ \begin{array}{l l l}
1 & 0 & 0 \\
0 & 1 & \gamma \\
0 & 0 & 1
\end{array} \right], \quad [\mathbf{b}] = \left[ \begin{array}{l c l}
1 & 0 & 0 \\
0 & 1 + \gamma^2 & \gamma \\
0 & \gamma & 1
\end{array} \right].
\end{equation}
Here $\gamma = \alpha R$ is the amount of shear at the radius $R$.

Substituting these values in the expressions \eqref{sigma = se+sv+pi}--\eqref{tauv value} for the total Cauchy stress, we obtain 
\begin{equation}
\sigma_{23} = \mu_e \gamma + \mu_v [\gamma - \gamma_v] - 2 n_e \gamma E_3^2 + 2 n_v \gamma^3 [-3 + 3 \gamma^2 + 2 \gamma^4] E_{e3}^2,
\end{equation}
where $\gamma_v$ is the corresponding shear strain for the viscous part of the deformation and $E_{e3} = E_3 - E_{v3}$.
The evolution equation \eqref{deform evolve} is reduced to
\begin{equation}
\frac{d \gamma_v}{dt} = \frac{1}{T_v} \left[ \gamma - \frac{1}{3} \left[ 3 + \left[ \gamma- \gamma_v \right]^2 \right] \gamma_v \right].
\end{equation}

Numerical results are obtained for following values of the material parameters
\begin{align}
T_v = 1 \, \mbox{s}, \quad \omega = 2 \, \mbox{s}^{-1},  \quad T_e = 2 \, \mbox{s}, \quad E_3 = 5\times 10^2 \, \mbox{V/m}.
\end{align}

\begin{figure}
 \begin{center}
  \begin{tabular}{c c}
   \includegraphics[scale=0.5]{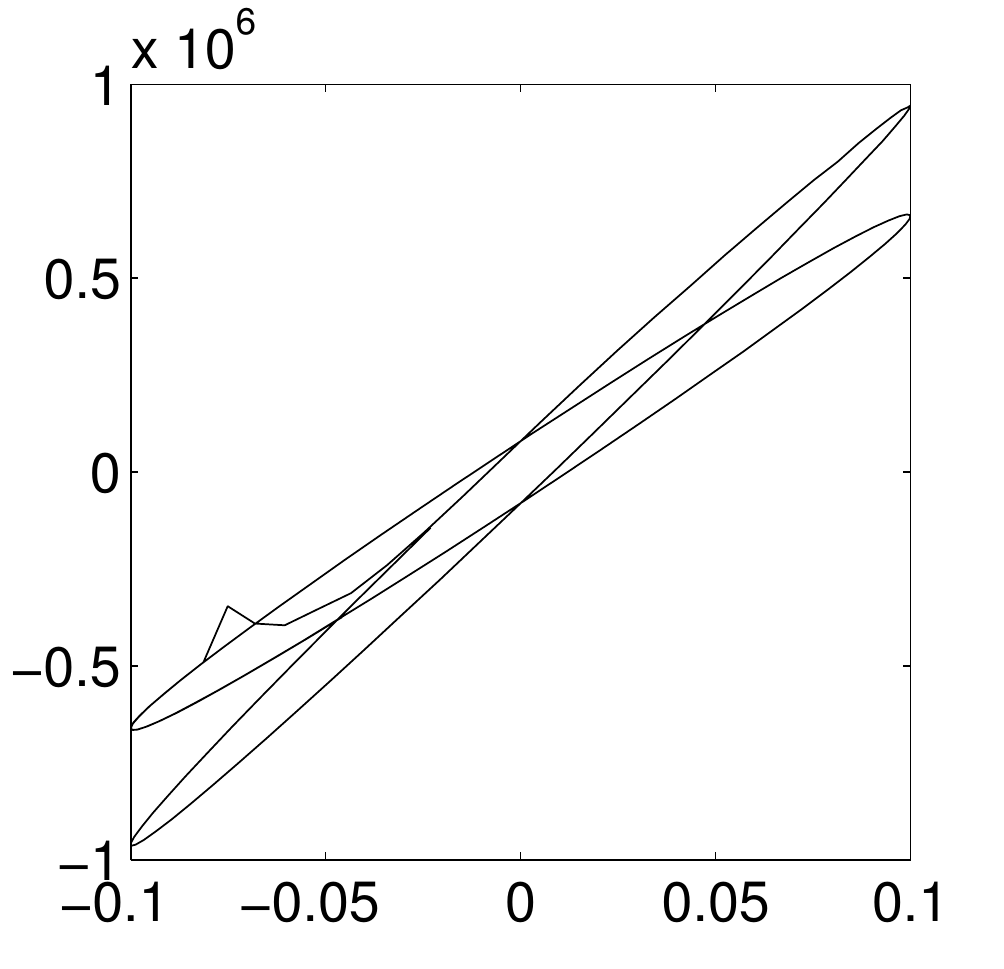} \hspace{20pt} & \hspace{20pt} \includegraphics[scale=0.5]{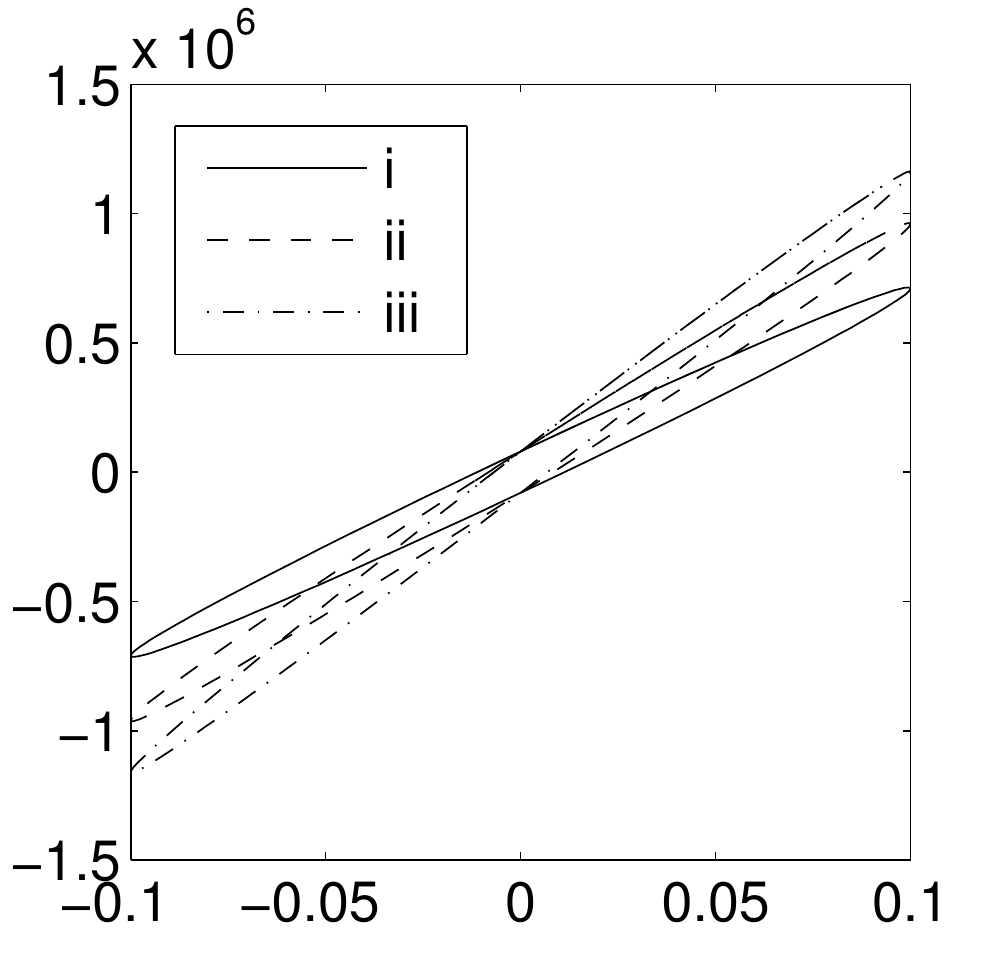} \\
\textbf{(a)} \hspace{35pt}& \hspace{35pt} \textbf{(b)}
  \end{tabular}
 \end{center}
\begin{picture}(0,0)(0,0)
 \put(140,30){$\gamma$} \put(335,30){$\gamma$}
\put(15, 110){$\sigma_{23}$} \put(210, 110){$\sigma_{23}$}
\put(130, 110){\scriptsize (i)} \put(115, 145){\scriptsize (ii)}
\put(45, 105){$\bullet$} \put(48, 108){\vector(0,-1){25}}
\end{picture}
\caption{Shear Cauchy stress $\sigma_{23}$ (N/m$^2$) vs the shear strain $\gamma$. (a) Switching on electric field during a dynamic shear experiment. The arrow shows the point where electric field is switched on resulting in a jump in $\sigma_{23}$. (i) $\mathbb{E}=0$, (ii) $\mathbb{E}_3 = 5 \times 10^2$ V/m. (b) Dependence on the electroelastic coupling parameter $n_e$: (i) $n_e = -1$ N/V$^2$, (ii) $n_e = -6$ N/V$^2$, (iii) $n_e = -10$ N/V$^2$. }
\label{fig: dynamic shear 1}
\end{figure}
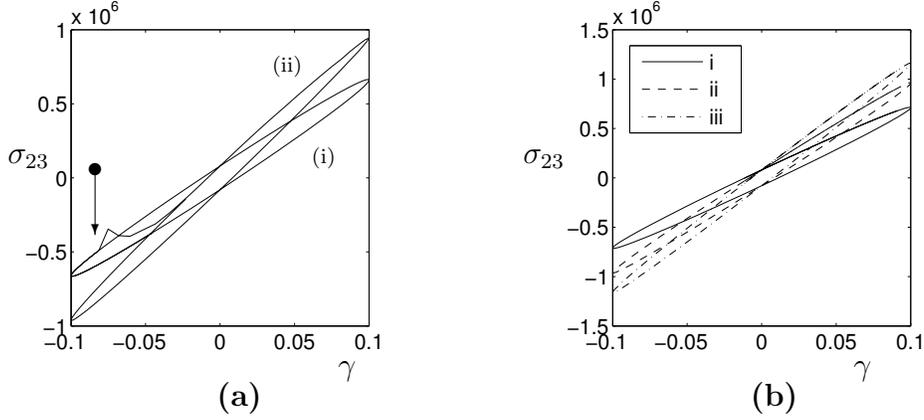

\begin{figure}
 \begin{center}
  \begin{tabular}{c c c}
   \hspace{-20pt} \includegraphics[scale=0.4]{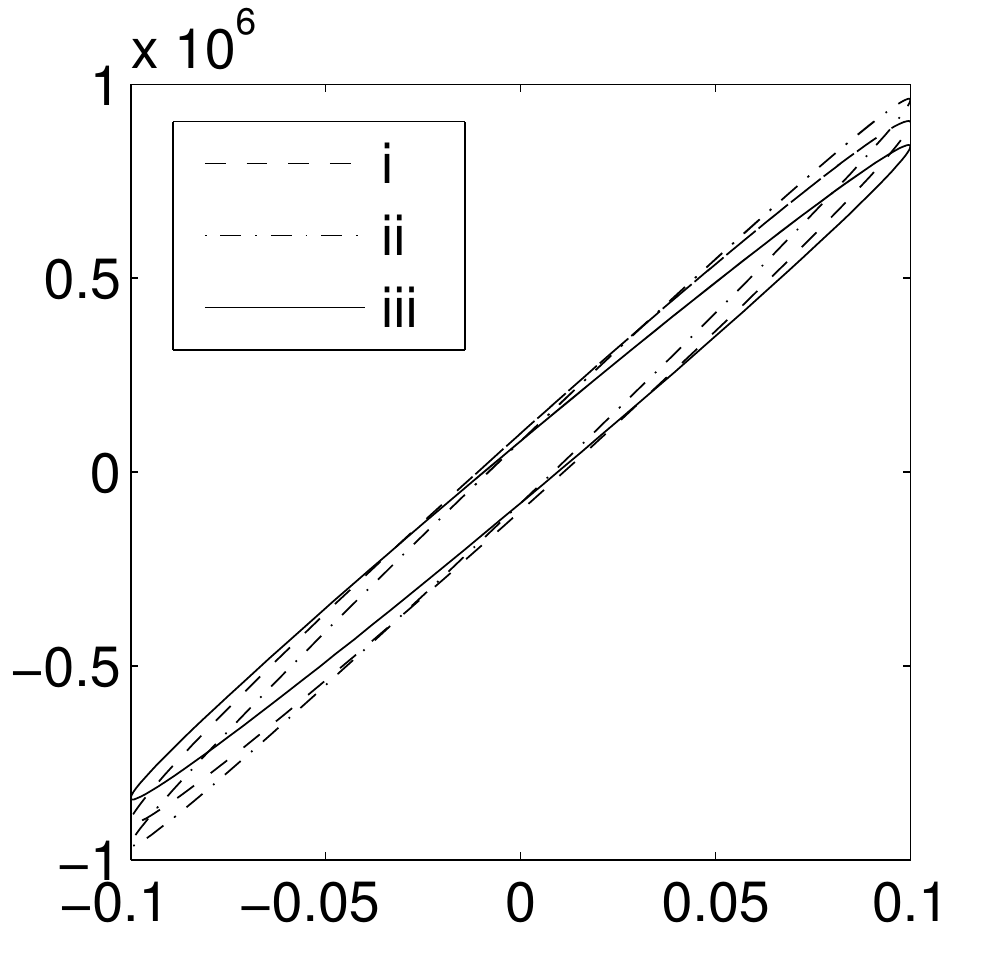} \hspace{10pt} & \hspace{10pt} \includegraphics[scale=0.4]{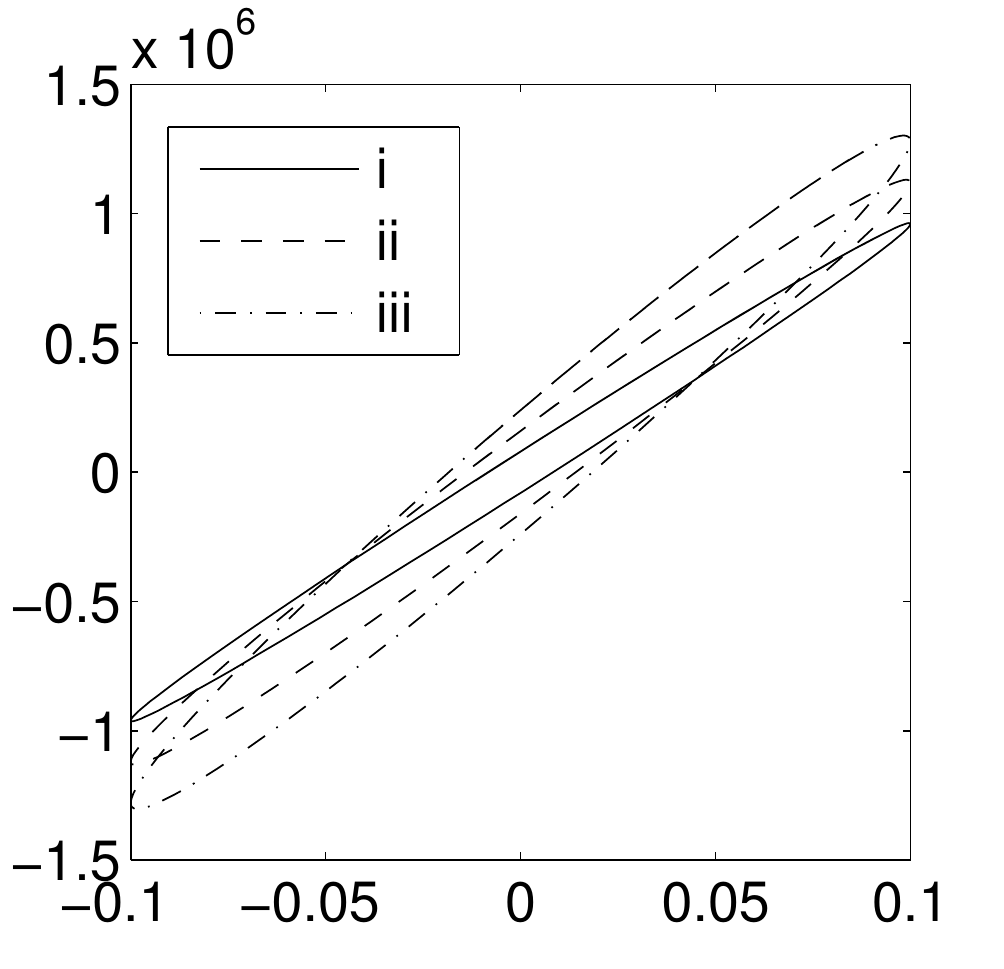} \hspace{10pt} & \hspace{10pt} \includegraphics[scale=0.4]{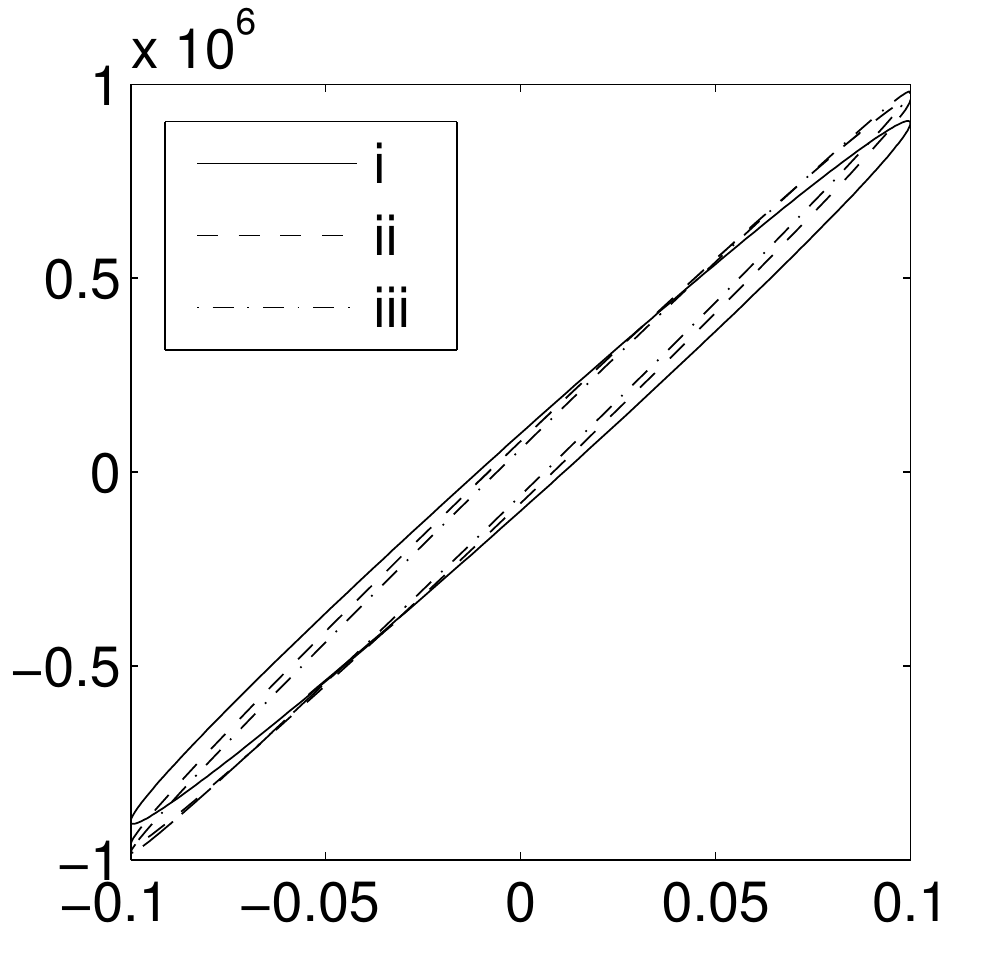} \\
 \hspace{-20pt} \textbf{(a)} \hspace{35pt}&  \textbf{(b)} & \hspace{20pt} \textbf{(c)}
  \end{tabular}
 \end{center}
\begin{picture}(0,0)(0,0)
\put(80,30){$\gamma$} \put(240,30){$\gamma$} \put(400,30){$\gamma$}
\put(-25, 120){$\sigma_{23}$} \put(135, 120){$\sigma_{23}$} \put(295, 120){$\sigma_{23}$}
\end{picture}
\caption{Dependence of the Lissajous plots on \textbf{(a)}~the parameter $T_v$: (i)~$T_v = 0.25$ s, (ii)~$T_v = 0.5$ s, (iii)~$T_v = 1$ s; \textbf{(b)}~the parameter $\mu_v$: (i)~$\mu_v = 2 \times 10^6$ MPa, (ii)~$\mu_v = 4 \times 10^6$ MPa, (iii)~$\mu_v = 6 \times 10^6$ MPa  and \textbf{(c)}~the oscillation frequency $\omega$: (i)~$\omega = 1$ s$^{-1}$, (ii)~$\omega = 2$ s$^{-1}$, (iii)~$\omega = 3$ s$^{-1}$.}
\label{fig: dynamic shear 2}
\end{figure}
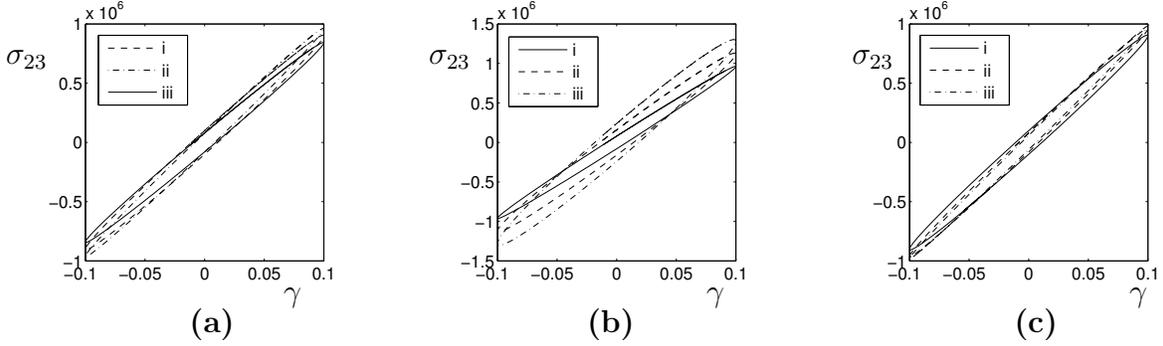

In the dynamic shear experiment, when measured over time, stress-strain curves form a hysteresis loop (also called the Lissajous plots) with the area enclosed by the curve related to the total energy loss per cycle. We show the dependence of such a curve on an applied electric field in Fig.~\ref{fig: dynamic shear 1}(a). The curves in this figure correspond to switching on of an electric field while the dynamic shear test is being performed -- curve (i) corresponds to $\mathbbm{e}=0$ while curve (ii) corresponds to $\mathbbm{e} = 500 $ V/m. As observed, the Cauchy stress ($\sigma_{23}$) gets a jump in its value on a sudden application of electric field and then evolves due to dissipation to converge to a steady-state region. The effect of the electroelastic coupling parameter $n_e$ on the dynamic shear experiment is demonstrated in Fig.~\ref{fig: dynamic shear 1}(b). Increasing the magnitude of $n_e$ causes a stronger coupling with the electric field, thus  increasing the slope of the curve which represents an increase in the effective shear modulus. 

Curves for different values of the parameter $T_v$ are shown in Fig.~\ref{fig: dynamic shear 2}(a). Increasing the value of $T_v$ results in lowering the slope and hence the effective shear modulus. It also results in a larger area enclosed in the curve thus increasing the energy loss during each load cycle. Reverse happens in the case when the parameter $\mu_v$ is changed in Fig.~\ref{fig: dynamic shear 2}(b). Increasing the value of $\mu_v$ results in a higher slope curve but also more dissipation per load cycle.
Plot for different oscillation frequencies $\omega$ in the presence of an electric field is shown in Fig.~\ref{fig: dynamic shear 2}(c). Increasing the oscillation frequency increases the slope of the ellipse thereby increasing the  effective shear modulus.

Dynamic shear test performed in a rheometer is a very widely used experimental procedure to characterise viscoelastic materials.  The curves presented in Figs.~\ref{fig: dynamic shear 1} and \ref{fig: dynamic shear 2} show that the presented model is able to capture serveral aspects of the Lissajous curves and therefore should be able to fit the experimental data effectively, as and when it becomes available.

\subsection{Extension and inflation of a compressible hollow cylinder}

Let an infinitely long hollow cylinder of internal radius $A$ and external radius $B$ be inflated and stretched in the axial direction such that the new internal and external radii are $a$ and $b$, respectively. Stretch in the axial direction is uniform such that a point at an axial position $Z$ is now at the axial coordinate $z=\lambda_z Z$ in the deformed configuration. It is assumed that there is no variation of deformation with the $\Theta$ coordinate and thus the deformation and the deformation gradient tensor in the cylindrical polar coordinate system ($R, \Theta, Z$) are given as
\begin{equation}
r(R) = R+ u(R), \quad \theta = \Theta, \quad z = \lambda_z Z,
\end{equation}
\begin{equation}
 \left[ \mathbf{F} \right] = \left[ \begin{array}{l l l}
g & 0 & 0 \\[3mm]
0 & \displaystyle{\frac{r}{R}} & 0 \\[3mm]
0 & 0 & \lambda_z
\end{array} \right],
\end{equation}
with $g = \partial r/\partial R$.

Along with the imposed mechanical boundary conditions, a potential difference $\Delta \phi = \phi_b - \phi_a$ is applied across the radial direction. This results in the generation of an electric field that is coupled with the underlying deformation.
In order to obtain the exact deformation and the variation of the electric field along the radial direction, the following system of equations needs to be solved:
\begin{align}
\mbox{Div}\, \mathbb{D} = 0, \quad \mbox{Div}\, \left(\mathbf{SF}^t \right) = \mathbf{0}, \quad \mathbb{E} = - \mbox{Grad}\, \phi, \label{gov cylindrical} \\
r(A) = a, \quad r(B) = b, \quad \phi(A) = \phi_a, \quad \phi(B) = \phi_b. \label{bc cylindrical}
\end{align}

We consider a compressible material for this problem and thus the energy function \eqref{energy omega E} is generalised to
\begin{equation}
 \Omega_e = \frac{\mu_e}{2} \left[ I_1 - 3 \right] - \mu_e \, \mbox{ln}\, J + \frac{\lambda_e}{2} \left[ \mbox{ln} \, J \right]^2 + m_e I_4 + n_e I_5  ,
\end{equation}
while the energy for the viscous part is still given by \eqref{energy omega V}.

Due to the geometric symmetry of the cylinder and the existence of potential difference across the radial direction, the electric field vanishes in the axial and the circumferential directions. For the given energy function, the electric displacement in radial direction is given as
\begin{equation}
\mathbb{D}_R = - 2 \left[ m_e \mathbb{E}_R + n_e g^{-2} \mathbb{E}_R + m_v \mathbb{E}_{e(R)} + n_v g^4 \mathbb{E}_{e(R)}  \right],
\end{equation}
while the nominal stress $\mathbf{T} = \mathbf{SF}^t$ in the three principal directions is given as
\begin{align}
T_{Rr} = \mu_e g + \frac{1}{2g} \left[ -\mu_e + \lambda_e \, \mbox{ln} \, J \right] + \frac{\mu_v g}{C_{v(RR)}} - \frac{ 2 n_e \mathbb{E}_{R}^2}{g^3}  + 2 n_v \mathbb{E}_{e(R)}^2 g^5 ,\\
T_{\Theta \theta} = \mu_e \frac{r}{R} + \frac{R}{2r} \left[ -\mu_e + \lambda_e \, \mbox{ln} \, J \right] + \frac{\mu_v r}{R C_{v(\Theta \Theta)}}, \\
T_{Zz} = \mu_e \lambda_z +\frac{\lambda_z}{2} \left[ -\mu_e + \lambda_e \, \mbox{ln} \, J \right]+ \mu_v \frac{\lambda_z}{C_{v(ZZ)}}.
\end{align}
The three principal components in the radial, azimuthal and axial directions of the tensor $\mathbf{C}_v$ are denoted by $C_{v(RR)}, C_{v(\Theta \Theta)}$ and $C_{v(ZZ)}$, respectively, in the above equations.

The governing equations \eqref{gov cylindrical} are reduced to
\begin{equation}
 \frac{\partial \left(R \mathbb{D}_R \right)}{\partial R} = 0, \quad \frac{\partial \left( R T_{RR} \right) }{\partial R} = T_{\Theta \Theta}, \quad \mathbb{E}_{R} = -\frac{\partial \phi}{\partial R},
\end{equation}
which need to be solved along with the boundary conditions \eqref{bc cylindrical} for every time step. The values of $\mathbb{E}_e$ and $\mathbf{C}_v$ are updated at every time step using the solution for $r(R), \phi(R)$, and the evolution equations \eqref{elec evolve} and \eqref{deform evolve}. It is noted that none of the governing equations relevant to this problem depend on the axial stretch $\lambda_z$.

\subsubsection{Numerial calculations}
Numerical calculations are performed for the following values of the material parameters and physical quantities
\begin{align}
 \mu_e = 5 \times 10^6 \, \mbox{N/m}^2, \quad \lambda_e = 6.6667\times 10^6 \, \mbox{N/m}^2,   \nonumber \\
m_e = - m_v = - 10 \, \mbox{N}/ \mbox{V}^2, \quad n_e = - 0.06 \mbox{N}/ \text{V}^2, \quad  n_v =  6 \, \mbox{N}/ \mbox{V}^2.
\end{align}
Computations  are performed using a finite-difference technique implemented in the \texttt{dsolve} solver of Maple for solving BVPs. 

\begin{figure}
\begin{center}
\begin{tabular}{c c}
 \includegraphics[scale=0.3]{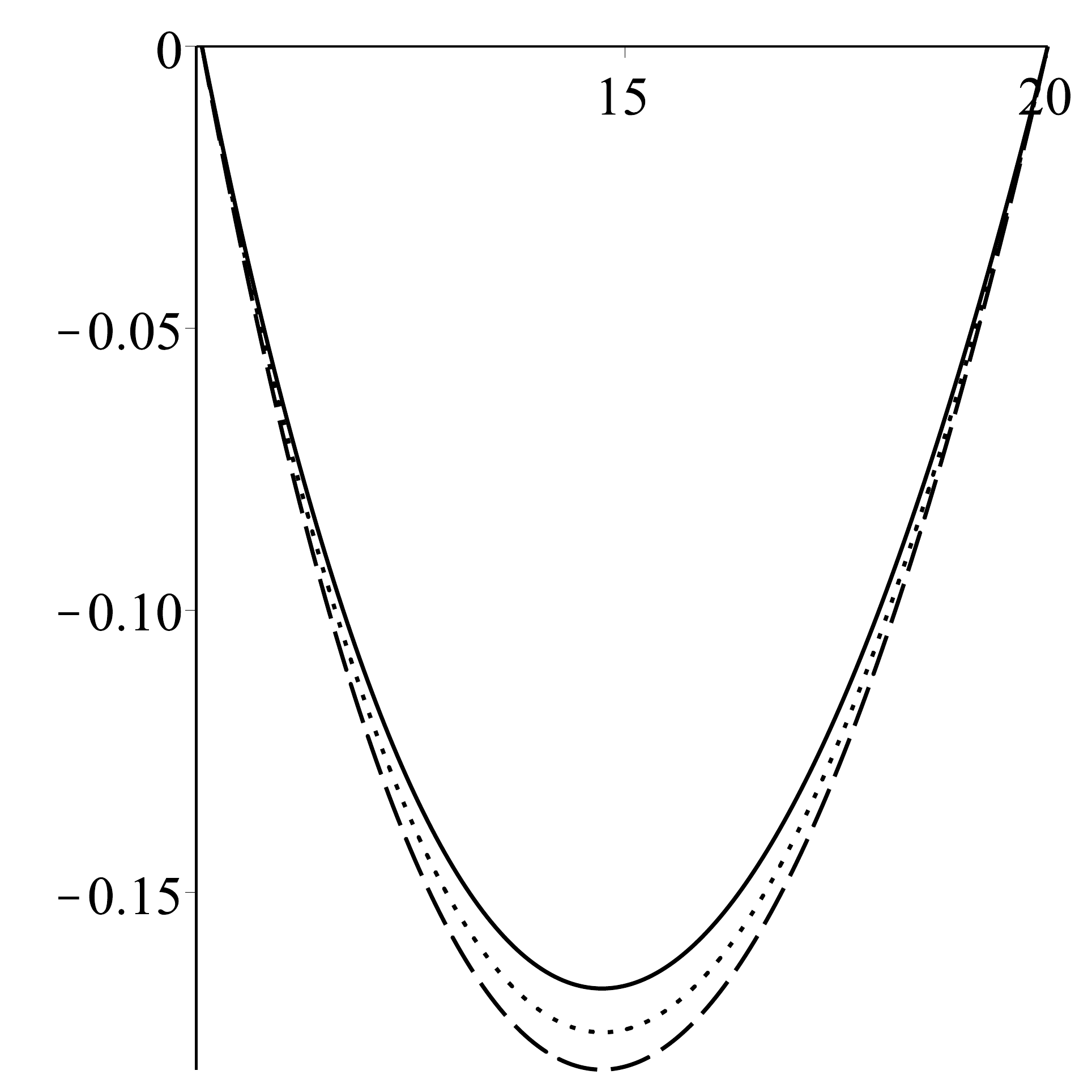} \hspace{10pt} & \hspace{10pt} \includegraphics[scale=0.3]{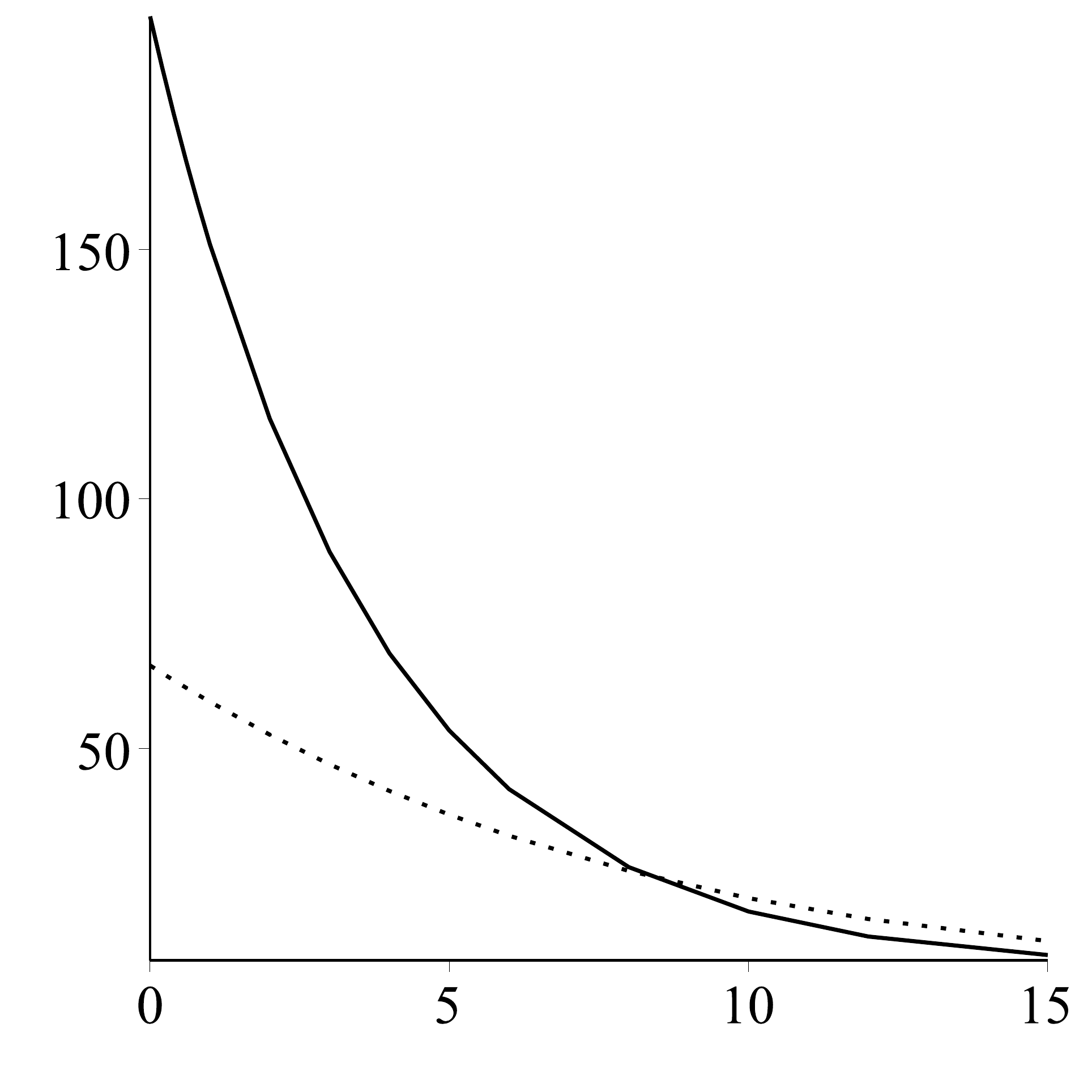} 
\end{tabular}
\end{center}
\begin{picture}(0,0)(0,0)
\put(120, 185){$R$} \put(345, 25){$t\,$}
\put(-5, 150){$\Delta \, R$}  \put(210, 150){$P$}  
\put(95, 5){\textbf{(a)}} \put(300, 5){\textbf{(b)}}
\put(140, 50){$t=10\,$s} \put(75, 70){$t=0$}
\put(255, 150){(i)} \put(255, 50){(ii)}
\end{picture}
\caption{(a) Displacement $\Delta R$ (mm) along the radius of the tube at time $t=\{ 0, 5, 10\}$ s. (b) Pressure (N/m$^2$) at (i) the internal boundary and (ii) the external boundary required to maintain the specified geometry.}
\label{fig: displ and stress}
\end{figure}
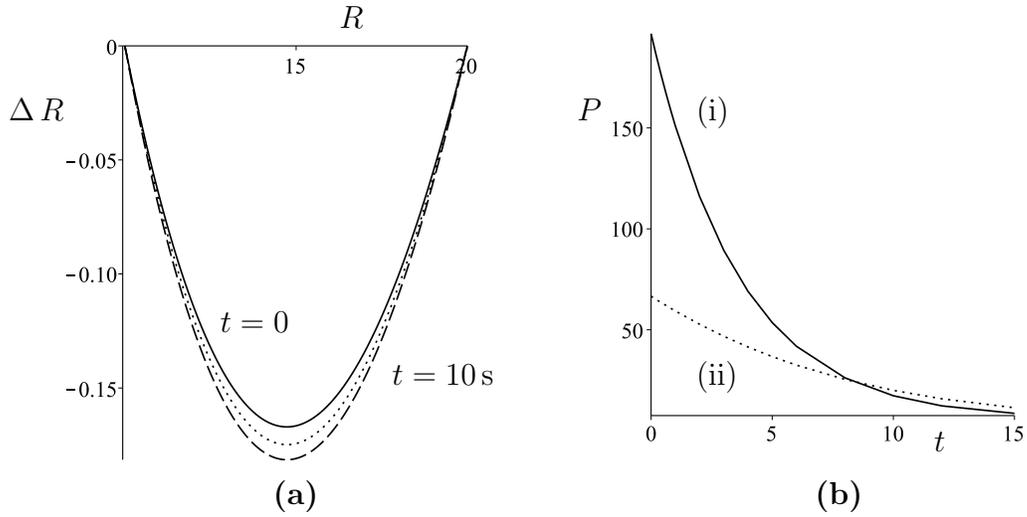


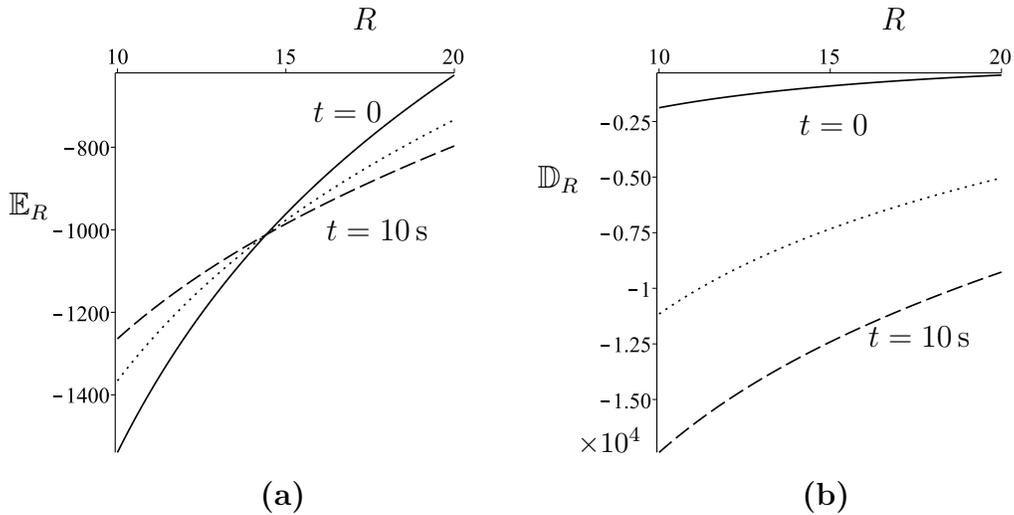
\begin{figure}
\begin{center}
\begin{tabular}{c c}
 \includegraphics[scale=0.3]{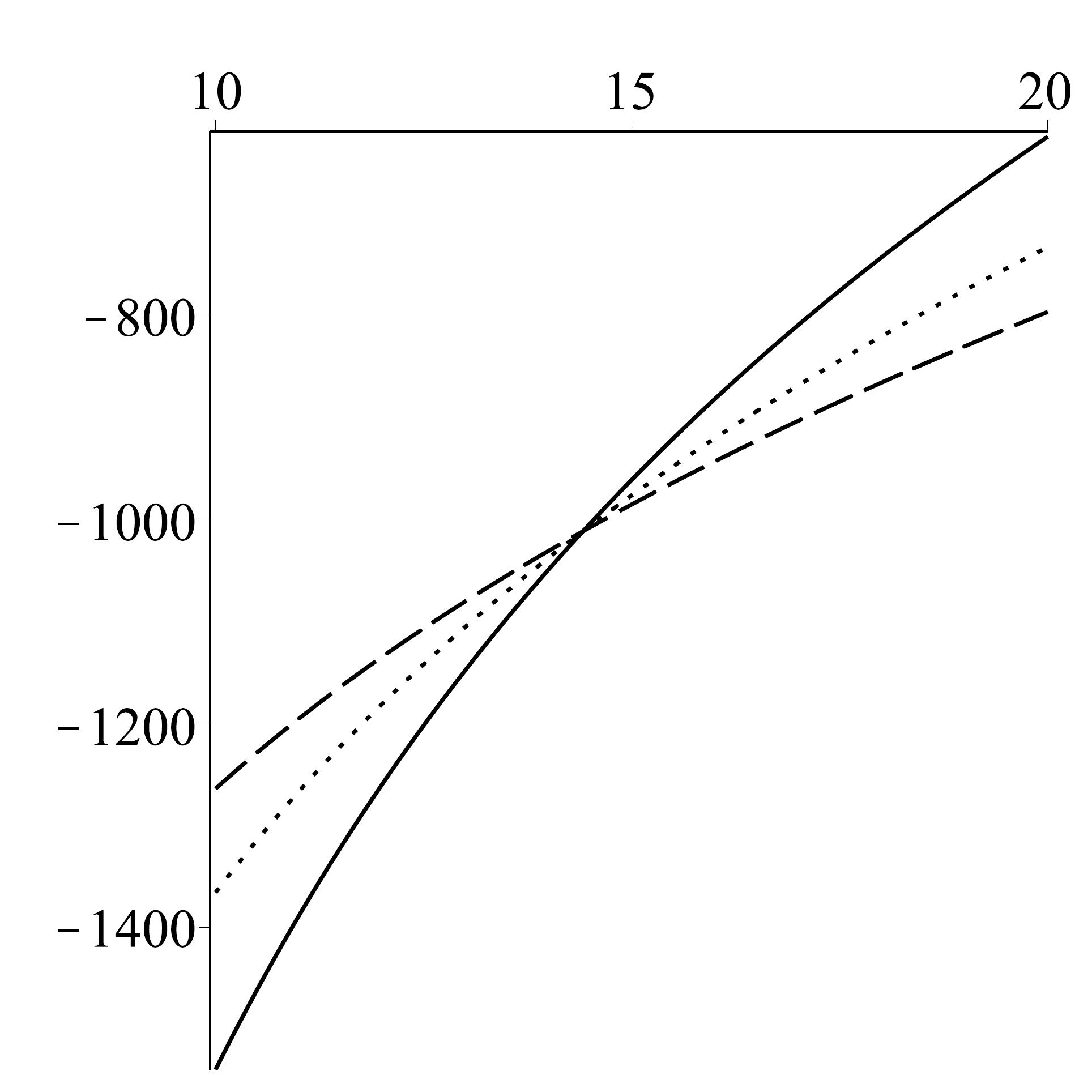} \hspace{10pt} & \hspace{10pt} \includegraphics[scale=0.3]{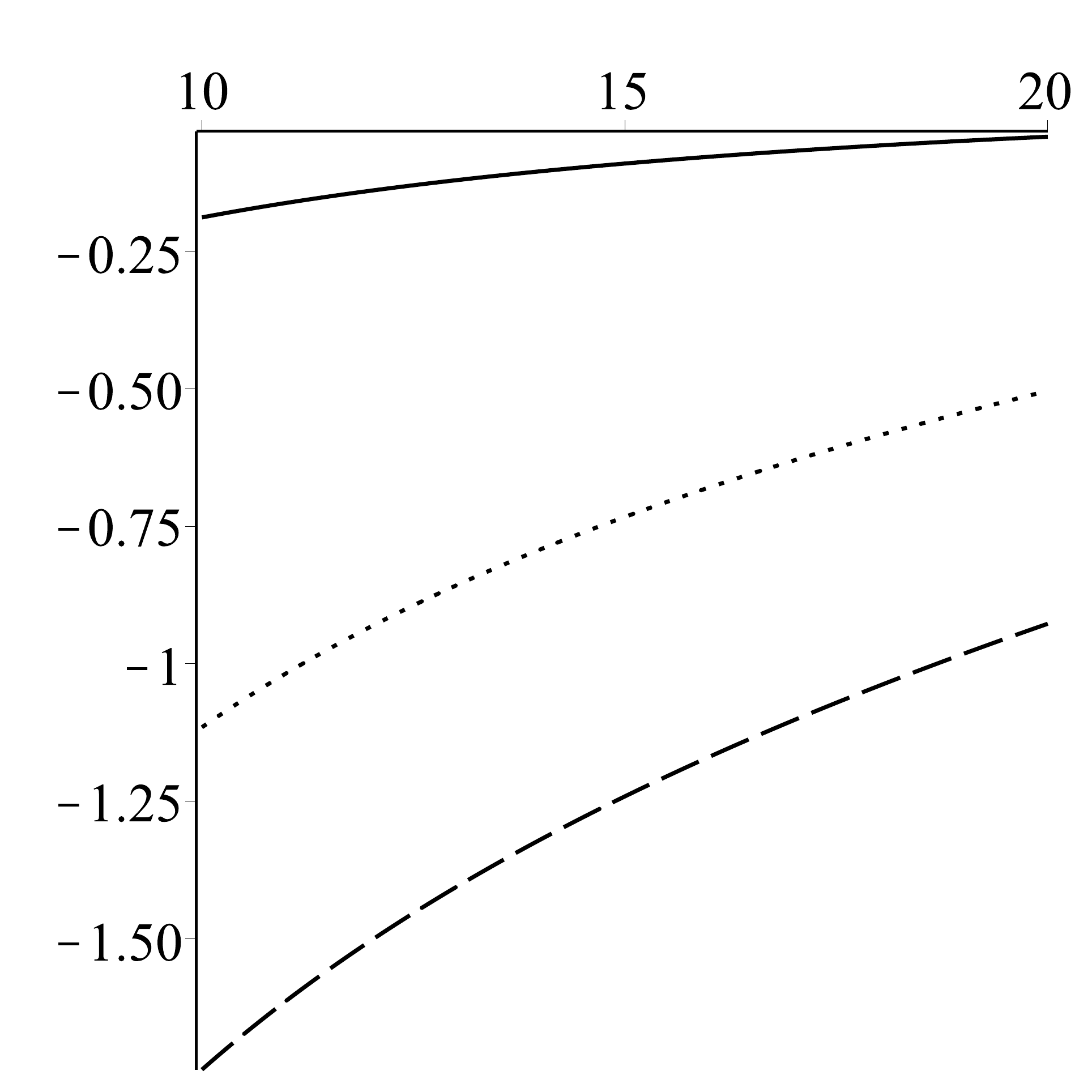}
\end{tabular}
\end{center}
\caption{(a) Radial electric field $\mathbb{E}_R$ (V/m) vs radius $R$ (mm). (b) Radial electric displacement $\mathbb{D}_R$ (N m$^{-1}$ V$^{-1}$) vs radius $R$ (mm) at times $t=\{ 0, 5, 10\}$~s.}
\begin{picture}(0,0)(0,0)
 \put(0, 160){$\mathbb{E}_R$} \put(130, 230){$R$}
\put(330, 230){$R$} \put(200, 170){$\mathbb{D}_R$}
\put(95, 50){\textbf{(a)}} \put(300, 50){\textbf{(b)}}
\put(120, 150){$t=10\,$s} \put(115, 195){$t=0$}
\put(299, 190){$t=0$} \put(325, 110){$t=10\,$s}
\put(215, 70){\small $\times 10^4$}
\end{picture}
\label{fig: E field and D field}
\end{figure}

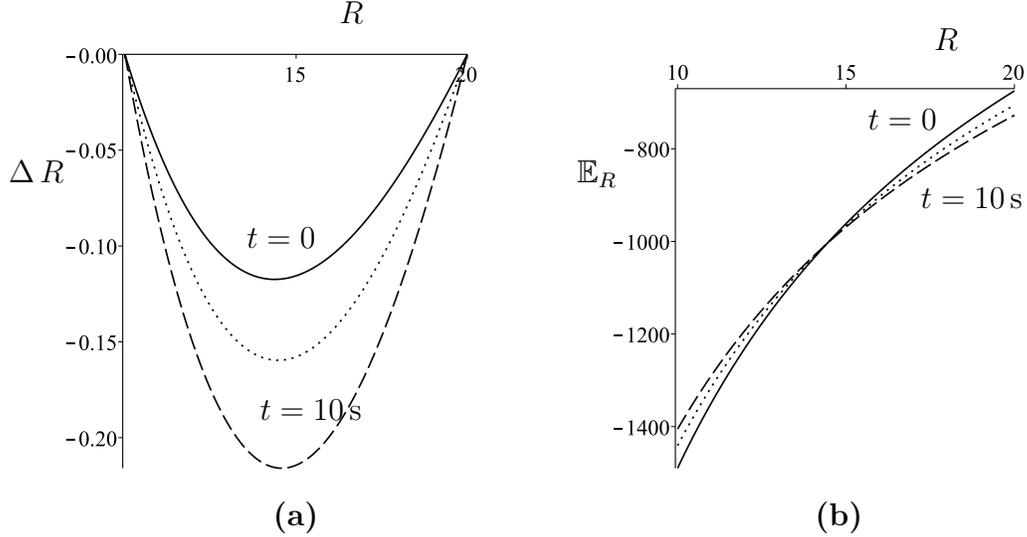
\begin{figure}
\begin{center}
\begin{tabular}{c c}
 \includegraphics[scale=0.3]{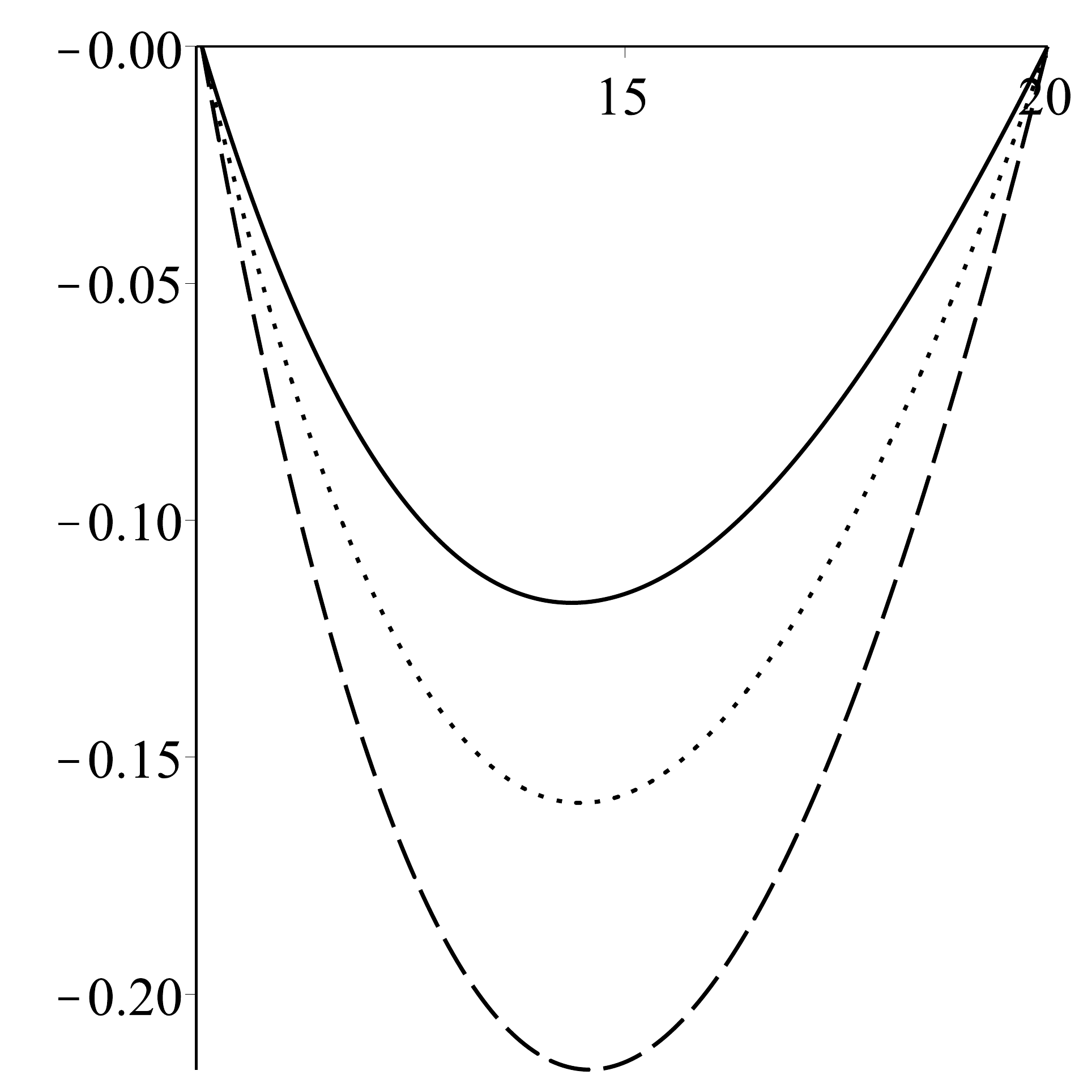} \hspace{10pt} & \hspace{10pt} \includegraphics[scale=0.3]{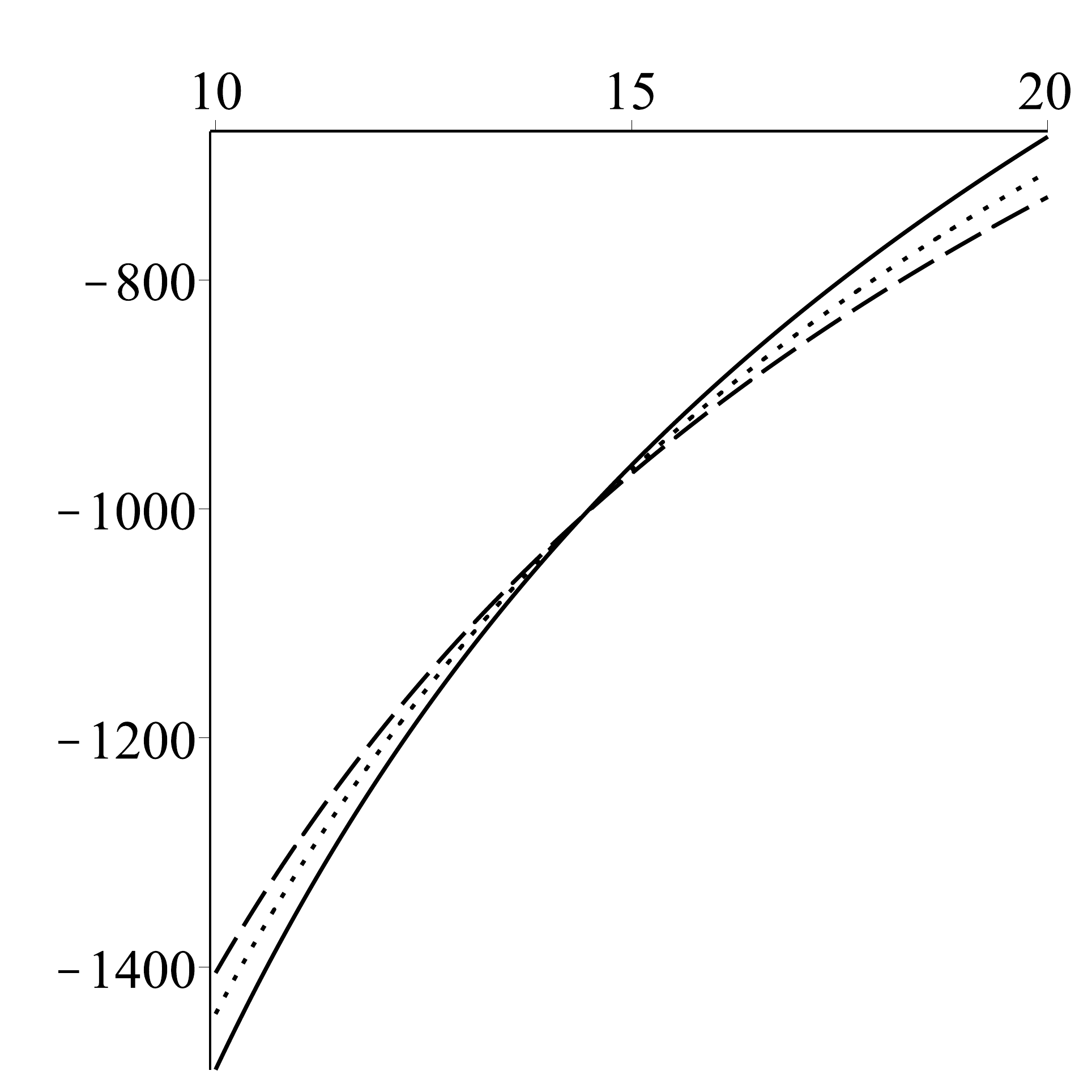}
\end{tabular}
\end{center}
\begin{picture}(0,0)(0,0)
\put(120, 190){$R \,$} \put(-5, 130){$\Delta \, R$} 
\put(95, 0){\textbf{(a)}} \put(300, 0){\textbf{(b)}}
\put(85, 105){$t=0$} \put(90, 40){$t= 10 \,$s}
\put(345, 180){$R\,$} \put(210, 130){$\mathbb{E}_R$} 
\put(320, 150){$t=0$} \put(340, 120){$t=10\,$s} 
\end{picture}
\caption{Displacement $\Delta R$ (mm) and the electric field $\mathbb{E}_R$ (V/m) along the radius of the tube at $t= \{0, 5, 10\}$ s  for $m_v = 5$ N/V$^2$.}
\label{fig: mv = 5}
\end{figure}
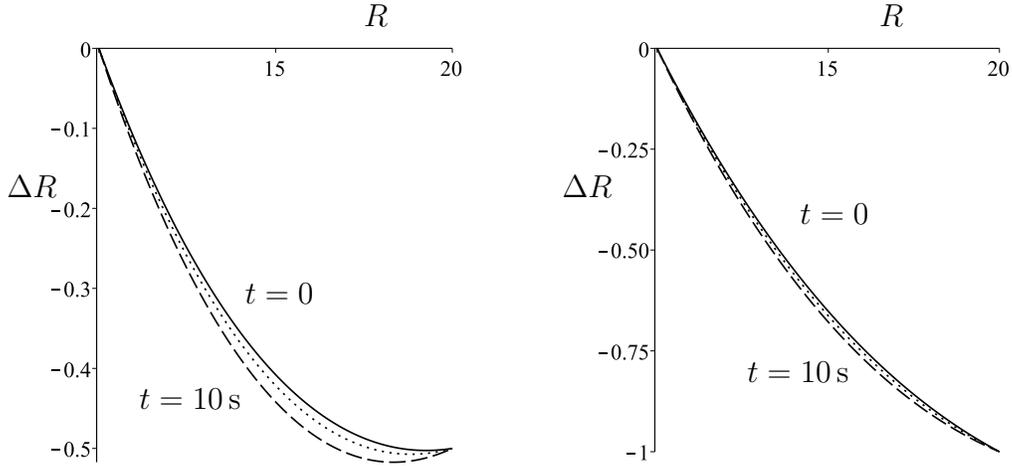
\begin{figure}
\begin{center}
\begin{tabular}{c c}
 \includegraphics[scale=0.3]{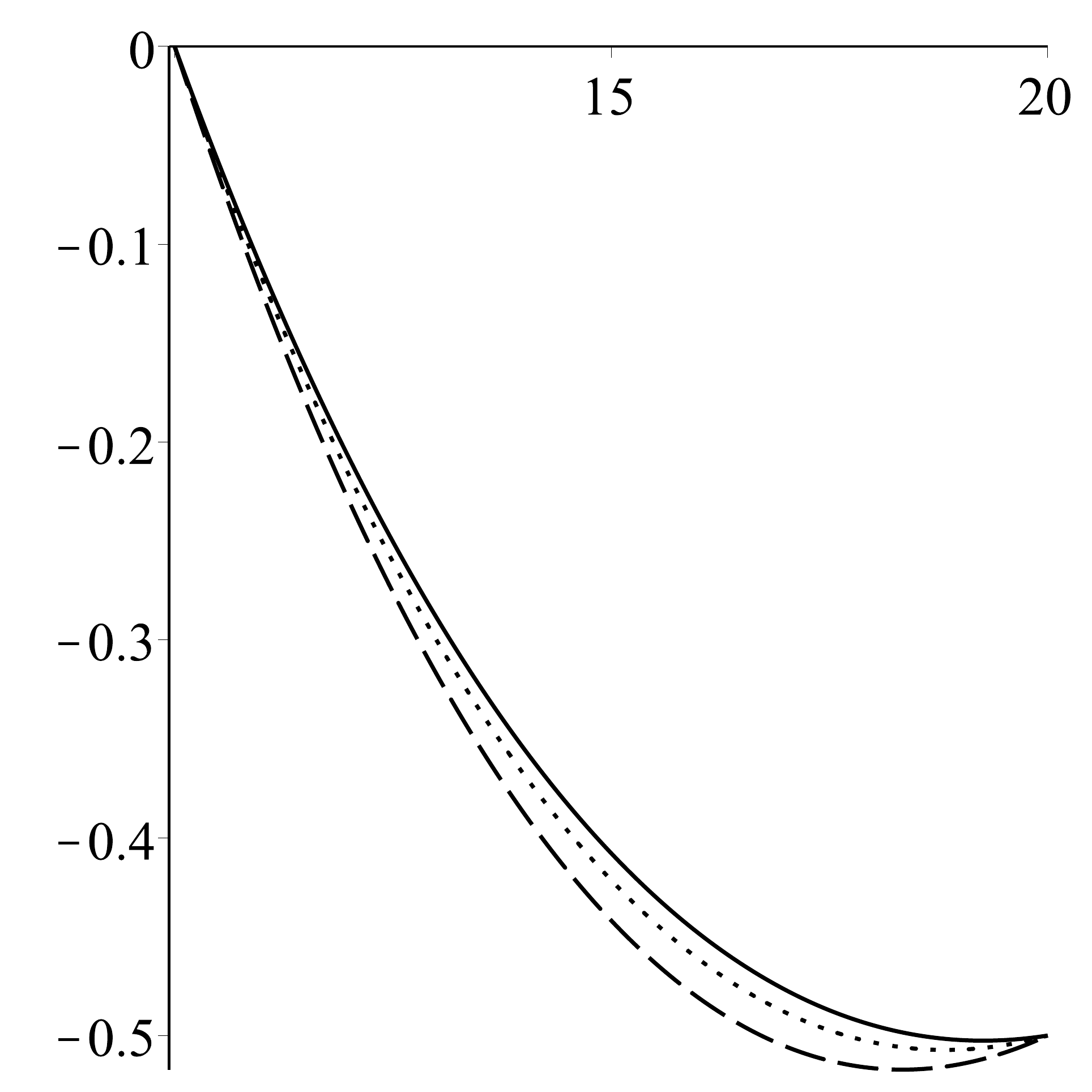} \hspace{10pt} & \hspace{10pt} \includegraphics[scale=0.3]{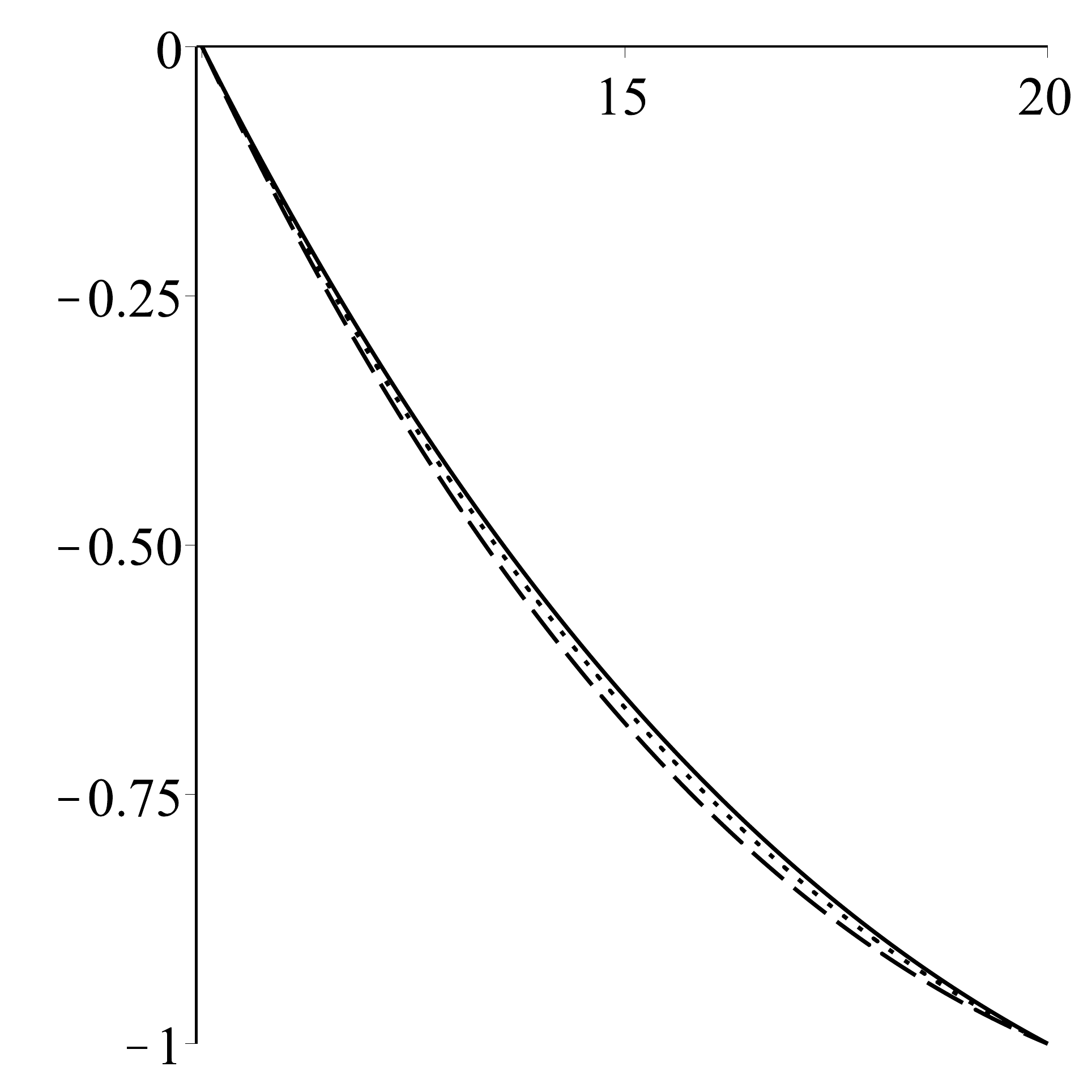}
\end{tabular}
\end{center}
\caption{Displacement $\Delta R$ (mm) vs radius $R$ for at $t= \{0, 5, 10\}$ s for (a)~$b = 19.5$ mm, (b)~$b = 19$ mm.}
\begin{picture}(0,0)(0,0)
\put(330,235){$R$} \put(210, 170){$\Delta R$}
\put(135, 235){$R$} \put(0, 170){$\Delta R$}
\put(50, 90){$t=10\,$s} \put(90, 130){$t= 0$}
\put(300, 160){$t=0$} \put(280, 100){$t= 10\,$s}
\end{picture}
\label{fig: delta R diff Rout}
\end{figure}

In order to understand the effects of the electric dissipation process, we neglect the mechanical viscosity for the time being. At time $t=0$, a potential difference of $\Delta \phi = 10$ V is applied across the internal and external surface of a tube with the internal and external radii given by $A = 10$ mm and $B = 20$ mm, respectively.
The resulting values of displacement $\Delta R= r-R$ as a function of radius $R$, electric field $\mathbb{E}$, and the electric displacement $\mathbb{D}$ are plotted with time in Figs.~\ref{fig: displ and stress}--\ref{fig: delta R diff Rout}.

In the first case, the internal and external boundaries of the tube are fixed such that $a=A$ and $b=B$.
The curves in Fig.~\ref{fig: displ and stress}(a) correspond to the displacement of points at different instants of time. Due to evolution of the `viscous' electric field $\mathbb{E}_v$ the displacement changes, starting from a small initial value it attains a larger equilibrium value with time. 
In Fig.~\ref{fig: displ and stress}(b), we plot the evolution with time of the internal and external pressures required to maintain the specified boundary conditions.

For the same problem, variation of electric field along the radius of the tube is plotted in Fig.~\ref{fig: E field and D field}(a) at different values of time. It is observed that the spatial gradient of the electric field decreases with time. There also exists a value of radius (say $R=R_0$) at which the value of electric field remains constant and for $R>R_0$ the magnitude of electric field increases while for $R<R_0$ the magnitude of electric field decreases with time.
The electric displacement attains maximum value at the internal boundary and minimum value at the external boundary. Evolution of $\mathbb{E}_v$ causes $\mathbb{D}$ to increase uniformly with time.

The effect of electro-viscoelastic coupling parameter $m_v$ is shown in Fig.~\ref{fig: mv = 5} in which an smaller value of $m_v = 5$ Pa m$^2$/V$^2$ is taken while keeping other parameters and boundary conditions the same. The resulting mechanical displacements and electric fields are plotted in Figs.~\ref{fig: mv = 5}(a) and \ref{fig: mv = 5}(b), respectively. It is observed in this case that the peak value of deformation is higher than for $m_v = 10$, and also this value is approached faster in comparison to the previous case.
Similar observations are made corresponding to the electric field. The peak values attained are higher in magnitude with a higher gradient along the radius.

As a last illustration, the mechanical boundary conditions for the problem are changed keeping other conditions constant. The internal wall of the cylinder is held at the same position while the external is compressed and brought to a known fixed radius. We consider the displacement of the external wall by $0.5$ mm and $1$ mm (corresponding to $b=19.5$ mm and $b=19$ mm, respectively) and plot the displacement along the radius in Figs.~\ref{fig: delta R diff Rout} (a) and \ref{fig: delta R diff Rout}(b) respectively. In this case, as the prescribed mechanical deformation of the outer wall is increased, the overall variation of $\Delta R$ with $R$ tends to be linear. Moreover, the  evolution of $\mathbb{E}_v$ also tends to have a lower effect on changing the values of $\Delta R$ in the case of larger values of prescribed displacements. This happens since in this case mechanical effects dominate the electrostatic forces.

\bigskip
\noindent \textbf{Acknowledgements:} This work is supported by an ERC advanced investigator grant towards the project MOCOPOLY.

\appendix
\section{Fundamental equations}

In the following sections, we derive the governing equations and the constitutive laws of electroelasticity from the basic  principles of electrostatics following the approach of Mcmeeking and Landis \cite{McMeeking2005}.

\subsection{Balance laws}

\subsubsection{Linear and angular momentum balance}
Conservation of mass implies that for any volume $V$, the following relation must hold
\begin{equation} \label{conservation of mass}
\bdot{\overline{\int\limits_V \rho \, \text{d}V}} = 0 \quad \Rightarrow \bdot{\rho} + \rho \, \mbox{div}\, \mathbf{v} = 0,
\end{equation}
where $\rho(\mathbf{x},t)$ is the mass density of the material and $\mathbf{v}$ is the velocity of a point $\mathbf{x}$ at time $t$. A superposed dot implies the material time derivative.
The laws of balance of linear and angular momenta are given as
\begin{equation}
 \int\limits_V  \left[\mathbf{f}_m + \mathbbm{f}_e \right] \text{d}V + \int\limits_S \left[ \mathbf{t}_m + \mathbbm{t}_e \right] \text{d}S = \bdot{\overline{ \int\limits_V \rho \mathbf{v} \, \text{d$V$}}},
\end{equation}
\begin{equation}
\int\limits_V \mathbf{x} \times \left[ \mathbf{f}_m + \mathbbm{f}_e \right] \text{d}V + \int\limits_S \mathbf{x} \times \left[ \mathbf{t}_m + \mathbbm{t}_e \right] \text{d}S = \bdot{\overline{ \int\limits_V  \mathbf{x} \times \rho \mathbf{v} \, \text{d$V$}}}.
\end{equation}
Here, $\mathbf{f}_m$ and $\mathbbm{f}_e$ are, respectively, the mechanical and the electrical body force per unit mass, and $\mathbf{t}_m$ and $\mathbbm{t}_e$ are, respectively, the mechanical and the electrical surface tractions. The electric body force is considered to be a result of the electric field acting inside the material and can be derived from a Maxwell (or electrical) stress tensor $\boldsymbol{\sigma}_e$ such that
\begin{equation}
\mathbbm{f}_e = \mbox{div}\, \boldsymbol{\sigma}_e.
\end{equation}
Thus, the electrical traction on the surface $S$ is given by
\begin{equation} \label{bc cond 54}
\mathbbm{t}_e = \llbracket \boldsymbol{\sigma}_e^t \rrbracket \mathbf{n},
\end{equation}
$\mathbf{n}$ being the unit outward normal to the surface $S$.

For an infinitesimally small surface element, in order to balance the linear momentum the mechanical Cauchy stress $\boldsymbol{\sigma}_m$ must balance the mechanical and the electrical traction, thus giving
\begin{equation} \label{bc 51}
 \mathbf{t}_m + \mathbbm{t}_e - \boldsymbol{\sigma}_m^t \mathbf{n} = 0, \quad \Rightarrow \mathbf{t}_m = - \llbracket \boldsymbol{\sigma}_m^t + \boldsymbol{\sigma}_e^t \rrbracket \mathbf{n}.
\end{equation}

Using the above equations, the balance of linear and angular momenta are now given as
\begin{equation}
\mbox{div} \left( \boldsymbol{\sigma}_m + \boldsymbol{\sigma}_e \right) + \mathbf{f}_m = \rho \mathbf{a},
\end{equation}
and
\begin{equation} \label{total stress symmetric}
 \boldsymbol{\sigma}_m + \boldsymbol{\sigma}_e = \boldsymbol{\sigma}_m^t + \boldsymbol{\sigma}_e^t.
\end{equation}

\subsubsection{External work on the system}

Rate of work done by the agencies external to the system is given by, cf. \cite{McMeeking2005}
\begin{equation}
\bdot{W} = \int\limits_{\mathcal B_t} \mathbf{f}_m \cdot \mathbf{v} \, \text{d}V + \int\limits_{\partial \mathcal B_t} \mathbf{t}_m \cdot \mathbf{v} \, \text{d}S + \int\limits_{\mathcal B_t} \phi \; \bdot{\overline{ \left[ q\, \text{d}V \right]}} + \int\limits_{\partial \mathcal B_t} \phi \stackrel{\bdot{\phantom{ }}}{\overline{{\hat{q}dS}}}. \label{dw/dt first}
\end{equation}
Here the first two terms correspond to the work done by the mechanical body force and the surface tractions, respectively. The third term is the statement of the rate of electric work -- electric potential $\phi$ multiplied by the rate of increment of charge $q\, dV$. This is the work required in bringing an infinitesimal charge $q \, dV$ from infinity to the point where the potential is $\phi$. The fourth term is the rate of work corresponding to the surface charge density.

It can be shown that
\begin{equation}
\bdot{\overline{ \left[q \, \text{d}V \right]}} = \left[ \frac{\partial q}{ \partial t} + \mathbf{v} \cdot \mbox{grad}\, q + q \, \mbox{div}\, \mathbf{v} \right] \text{d}V,
\end{equation}
and
\begin{equation}
\bdot{\overline{ \left[ \llbracket \mathbbm{d} \cdot \mathbf{n} \rrbracket \text{d$S$} \right] }} = \left[ \Bigl\llbracket \bdot{\mathbbm{d}}  + \mathbbm{d} \, \mbox{div}\, \mathbf{v} - \left[ \mbox{grad}\, \mathbf{v} \right] \mathbbm{d} \Bigr\rrbracket \cdot \mathbf{n} \right] \text{d}S.
\end{equation}
In the second equation, use has been made of the Nanson's formula connecting material and spatial surface elements as $\mathbf{n} \, \text{d}S = J \mathbf{F}^{-t} \mathbf{N}\, \text{d}S_r$.

Substituting the above equations in \eqref{dw/dt first} and using the boundary condition \eqref{bc 51}, we get
\begin{align}
\bdot{W} = \int\limits_{\mathcal B_t}  \left[ \rho \mathbf{a} - \mbox{div} \left( \boldsymbol{\sigma}_m + \boldsymbol{\sigma}_e \right) \right] \cdot \mathbf{v} \, \text{d} V - \int\limits_{\partial \mathcal B_t}  \left[ \llbracket \boldsymbol{\sigma}_m^t + \boldsymbol{\sigma}_e^t \rrbracket \mathbf{n} \right] \cdot \mathbf{v} \, \text{d} S  \nonumber \\
+  \int\limits_{\mathcal B_t} \phi  \left[ \frac{\partial q}{ \partial t} + \mathbf{v} \cdot \mbox{grad}\, q + q \, \mbox{div}\, \mathbf{v} \right] \text{d}V \nonumber \\
+ \int\limits_{\partial \mathcal B_t} \phi \left[ \Bigl\llbracket \bdot{\mathbbm{d}}  + \mathbbm{d} \left[ \mbox{div}\, \mathbf{v} \right] - \left[ \mbox{grad}\, \mathbf{v} \right] \mathbbm{d} \Bigr\rrbracket \cdot \mathbf{n} \right] \text{d}S.
\end{align}

The two surface integrals can be converted to volume integrals using the divergence theorem to give
\begin{align}
\bdot{W} = \int\limits_{\mathcal B_t} \left[ \left[ \rho \mathbf{a} - \mbox{div} \left( \boldsymbol{\sigma}_m + \boldsymbol{\sigma}_e \right) \right] \cdot \mathbf{v} + \phi  \left[ \frac{\partial q}{ \partial t} + \mathbf{v} \cdot \mbox{grad}\, q + q \, \mbox{div}\, \mathbf{v} \right] \right. \nonumber \\
\left. - \mbox{div}\left( \phi \left[ \bdot{\mathbbm{d}}  + \mathbbm{d} \left[ \mbox{div}\, \mathbf{v} \right] - \left[ \mbox{grad}\, \mathbf{v} \right] \mathbbm{d} \right] \right) + \mbox{div}\, \left( \left[ \boldsymbol{\sigma}_m + \boldsymbol{\sigma}_e \right] \mathbf{v} \right) \right] \text{d}V.
\end{align}

On substituting the relations \eqref{gov euler 1}$_2$, \eqref{E potential}, and \eqref{conservation of mass} in the above equation, and a subsequent rearrangement of terms, we obtain
\begin{equation} \label{dW/dt expression}
 \bdot{W} = \int\limits_{\mathcal B_t} \left[ \rho \mathbf{a} \cdot \mathbf{v} + \left[ \boldsymbol{\sigma}_m + \boldsymbol{\sigma}_e - \mathbbm{d} \otimes \mathbbm{e} \right]: \mbox{grad}\, \mathbf{v} + \mathbbm{e} \cdot \bdot{\mathbbm{d}} + \left[ \mbox{div}\, \mathbf{v} \right] \mathbbm{e} \cdot \mathbbm{d} \right] \text{d}v.
\end{equation}

We now refer to the expression $\left[\boldsymbol{\sigma}_m + \boldsymbol{\sigma}_e \right]$ as the total stress tensor  which is a symmetric tensor (from equation \eqref{total stress symmetric}) including both the mechanical and the electric effects. Thus a total symmetric Piola--Kirchhoff stress can now be defined as $\mathbf{S}^{\text{tot}} = J \mathbf{F}^{-1} \left[ \boldsymbol{\sigma}_m + \boldsymbol{\sigma}_e \right] \mathbf{F}^{-t}$ using which the above statement for work rate is written in the referential form as
\begin{equation}
\bdot{W} = \int\limits_{\mathcal B_r} \left[ \rho_r \mathbf{v} \cdot \mathbf{a} + \frac{1}{2} \mathbf{S}^{\text{tot}} : \bdot{\mathbf{C}} + \mathbb{E} \cdot \bdot{\mathbb{D}} \right] \text{d}V_r.
\end{equation}

\subsection{Thermodynamics and constitutive relations}

By the first law of thermodynamics, balance of energy is stated as a relation between work done on the system, internal energy of the system and the heat transferred to the system as
\begin{equation} \label{FLT statement}
\bdot{U} = \bdot{W} + \bdot{Q},
\end{equation}
where $U$ represents the energy stored in the system and $\bdot{Q}$ is the rate of heat transfer to the system. Specifically, they can be written as
\begin{equation} \label{dU/dt expression}
\bdot{U}  = \frac{d}{dt} \int\limits_{\mathcal B_t} \rho u \, \text{d}v + \frac{d}{dt}  \int\limits_{\mathcal B_t} \frac{1}{2} \varepsilon_0 \mathbbm{e} \cdot \mathbbm{e}\, \text{d}v  + \frac{d}{dt}  \int\limits_{\mathcal B_t} \frac{1}{2} \rho \mathbf{v} \cdot \mathbf{v} \, \text{d}v,
\end{equation}
\begin{equation} \label{dQ/dt expression}
\bdot{Q} = \int\limits_{\mathcal B_t} \rho r  \, \text{d}v  - \int\limits_{\partial \mathcal B_t} \mathbf{q} \cdot \mathbf{n} \, \text{d}s,
\end{equation}
where $u$ is the  internal energy per unit mass and $r$ is the rate of heat transfer per unit mass.

In referential form, they give
\begin{equation}
\bdot{U} = \frac{d}{dt} \int\limits_{\mathcal B_0} \left[ \rho_r u + \frac{1}{2} \varepsilon_0 J \mathbb{E} \cdot \left[ \mathbf{C}^{-1} \mathbb{E} \right] + \frac{1}{2} \rho_r \mathbf{v} \cdot \mathbf{v} \right] \text{d}V_r,
\end{equation}
\begin{equation}
\bdot{Q} = \int\limits_{\mathcal B_0} \left[ \rho_r r - \mbox{Div}\, \mathbf{Q} \right] \, \text{d}V_r ,
\end{equation}
where $\mathbf{Q} = J \mathbf{F}^{-1} \mathbf{q}$ is the heat flux in the reference configuration.

Thus the statement of first law \eqref{FLT statement} is given in referential form as
\begin{align}
\rho_r \bdot{u} - \frac{1}{2} \mathbf{S}^{\text{tot}} : \bdot{\mathbf{C}} - \mathbb{E} \cdot \bdot{\mathbb{P}} + \frac{1}{2} \bdot{\mathbb{E}} \cdot \left[ \varepsilon_0 J \mathbf{C}^{-1} \mathbb{E} \right]  - \frac{1}{2} \mathbb{E} \cdot \left[ \varepsilon_0 \bdot{\overline{J \mathbf{C}^{-1} \mathbb{E}}} \right] \nonumber \\
 - \rho_r r + \mbox{Div}\, \mathbf{Q} = 0.
\end{align}

We now define a total energy function similar to the one given by Dorfmann and Ogden \cite{Dorfmann2006} to take into account the energy of electric field
\begin{equation}
 \psi = \rho_r u - \frac{1}{2} \varepsilon_0 J \mathbb{E} \cdot \left[ \mathbf{C}^{-1} \mathbb{E} \right].
\end{equation}
On substituting $\psi$ into the equation above, we obtain
\begin{equation}
 \bdot{\psi} - \frac{1}{2} \mathbf{S}^{\text{tot}} : \bdot{\mathbf{C}} - \mathbb{E} \cdot \bdot{\mathbb{P}} - \rho_r r + \mbox{Div}\, \mathbf{Q}+ \varepsilon_0 J \left[ \mathbf{C}^{-1} \mathbb{E} \right] \cdot \bdot{\mathbb{E}} = 0.
\end{equation}

On using a Legendre's transformation to change the dependent variable from $\mathbb{P}$ to $\mathbb{E}$
\begin{equation}
\varphi = \psi - \mathbb{E} \cdot \mathbb{P},
\end{equation}
we eventually obtain the first law in the desired form
\begin{equation} \label{FLT final}
\bdot{\varphi} - \frac{1}{2} \mathbf{S}^{\text{tot}} : \bdot{\mathbf{C}} + \mathbb{D} \cdot \bdot{\mathbb{E}} - \rho_r r + \mbox{Div}\, \mathbf{Q} = 0.
\end{equation}

Let $s$ be the entropy density per unit mass, then the second law of thermodynamics is given as
\begin{equation}
\frac{d}{dt}  \int\limits_{\mathcal B_t} \rho s \, \text{d}V  \ge \int\limits_{\mathcal B_t} \frac{\rho r}{\vartheta} \text{d}V - \int\limits_{\partial \mathcal B_t} \frac{1}{\vartheta} \mathbf{q} \cdot \mathbf{n} \, \text{d}S,
\end{equation}
which when written in terms of referential quantities, gives
\begin{equation}
\frac{d}{dt}  \int\limits_{\mathcal B_{\mbox{\scriptsize r}}} \rho_r s \, \text{d}V  \ge \int\limits_{\mathcal B_r} \frac{\rho_r r}{\vartheta} \text{d}V_r - \int\limits_{\partial \mathcal B_r} \frac{1}{\vartheta} \mathbf{Q} \cdot \mathbf{N} \, \text{d}S_r.
\end{equation}

On using the divergence theorem on the last term, we write the above inequality in local form to obtain
\begin{equation} 
 \rho_r \bdot{s} \ge \frac{\rho_r r}{\vartheta} - \frac{1}{\vartheta} \mbox{Div}\, \mathbf{Q} + \frac{1}{\vartheta^2} \mathbf{Q} \cdot \mbox{Grad}\, \vartheta.
\end{equation}

Substituting the statement of first law \eqref{FLT final} into the inequality above gives
\begin{equation} \label{SLT 38}
- \bdot{\varphi} + \frac{1}{2} \mathbf{S}^{\text{tot}} : \bdot{\mathbf{C}} - \mathbb{D} \cdot \bdot{\mathbb{E}} - \rho_r \vartheta \bdot{s} -  \frac{1}{\vartheta} \mathbf{Q} \cdot \mbox{Grad}\, \vartheta \ge 0.
\end{equation}

A final Legendre's transformation is now performed to change the dependency from $s$ to $\vartheta$
\begin{equation}
 \Omega = \varphi - \rho_r \vartheta s.
\end{equation}
Substitution of the above defined $\Omega$ into inequality \eqref{SLT 38} gives
\begin{equation} \label{SLT last appendix}
 - \bdot{\Omega} + \frac{1}{2} \mathbf{S}^{\text{tot}} : \bdot{\mathbf{C}} - \mathbb{D} \cdot \bdot{\mathbb{E}} - \rho_r s \bdot{\vartheta} - \frac{1}{\vartheta} \mathbf{Q} \cdot \mbox{Grad}\, \vartheta \ge 0.
\end{equation}

We now note that for an incompressible material, the constraint $J=1$ gives
\begin{equation}
 \bdot{\overline{ J^2}} = \mathbf{C}^{-1} : \bdot{\mathbf{C}} =0.
\end{equation}
Thus a scalar multiple of this zero term can be added to the above inequality without changing its meaning.


\begin{thebibliography}{10}

\bibitem{Pao1978}
Y.~H. Pao, ``{Electromagnetic forces in deformable continua},'' in {\em
  Mechanics Today, Vol. 4} (S.~Nemat-Nasser, ed.), pp.~209--305, Oxford
  University Press, 1978.

\bibitem{Eringen1990}
A.~C. Eringen and G.~A. Maugin, {\em {Electrodynamics of Continua, Vol. 1}}.
\newblock Springer-Verlag, 1990.

\bibitem{O'Halloran2008}
A.~O'Halloran, F.~O'Malley, and P.~McHugh, ``{A review on dielectric elastomer
  actuators, technology, applications, and challenges},'' {\em Journal of
  Applied Physics}, vol.~104, no.~7, pp.~71101--71110, 2008.

\bibitem{Wingert2006}
A.~Wingert, M.~D. Lichter, and S.~Dubowsky, ``{On the design of large
  degree-of-freedom digital mechatronic devices based on bistable dielectric
  elastomer actuators},'' {\em IEEE/ASME Transactions on Mechatronics},
  vol.~11, no.~4, pp.~448--456, 2006.

\bibitem{Bowers2011}
A.~E. Bowers, J.~M. Rossiter, P.~J. Walters, and I.~A. Ieropoulos,
  ``{Dielectric elastomer pump for artificial organisms},'' in {\em Proceedings
  of SPIE - EAPAD}, p.~797629, 2011.

\bibitem{Heydt2008}
R.~Heydt, R.~Kornbluh, J.~Eckerle, and R.~Pelrine, ``{Dielectric elastomer
  loudspeakers},'' in {\em Dielectric elastomers as electromechanical
  transducers: fundamentals, materials, devices, models and applications of a
  emerging electroactive polymer technology} (F.~Carpi, D.~DeRossi, and
  R.~Kornbluh, eds.), pp.~313--320, Elsevier Science Ltd., Amsterdam, 2008.

\bibitem{Zhang2006}
R.~Zhang, A.~Kunz, P.~Lochmatter, and G.~Kovacs, ``{Dielectric elastomer spring
  roll actuators for a portable force feedback device},'' in {\em 14th
  Symposium on Haptic Interfaces for Virtual Environment and Teleoperator
  Systems}, pp.~347--353, 2006.

\bibitem{Ozsecen2010}
M.~Y. Ozsecen, M.~Sivak, and C.~Mavroidis, ``{Haptic interfaces using
  dielectric electroactive polymers},'' in {\em Proceedings of SPIE - Sensors
  and Smart Structures Technologies for Civil, Mechanical, and Aerospace
  Systems} (M.~Tomizuka, C.~B. Yun, V.~Giurgiutiu, and J.~P. Lynch, eds.),
  p.~7647, 2010.

\bibitem{Pelrine2001}
R.~Pelrine, R.~Kornbluh, J.~Eckerle, P.~Jeuck, S.~Oh, Q.~Pei, and S.~Stanford,
  ``{Dielectric elastomers: generator mode fundamentals and applications},''
  {\em Proceedings of SPIE - Smart Structures and Materials}, vol.~4329,
  pp.~148--156, 2001.

\bibitem{McKay2011}
T.~G. McKay, B.~M. O'Brien, E.~P. Calius, and I.~A. Anderson, ``{Soft
  generators using dielectric elastomers},'' {\em Applied Physics Letters},
  vol.~98, p.~142903, 2011.

\bibitem{Anderson2011}
I.~A. Anderson, T.~C.~H. Tse, T.~Inamura, B.~M. O'Brien, T.~McKay, and
  T.~Gisby, ``{A soft and dexterous motor},'' {\em Applied Physics Letters},
  vol.~98, p.~123704, 2011.

\bibitem{Michel2008}
S.~Michel, A.~Bormann, C.~Jordi, and E.~Fink, ``{Feasibility studies for a
  bionic propulsion system of a blimp based on dielectric elastomers},'' {\em
  Proceedings of SPIE - EAPAD}, vol.~4332, pp.~1--15, 2008.

\bibitem{Toth2002}
L.~A. Toth and A.~A. Goldenberg, ``{Control system design for a dielectric
  elastomer actuator: the sensory subsystem},'' {\em Proceedings of SPIE -
  Smart Structures and Materials}, vol.~4695, pp.~323--334, 2002.

\bibitem{Son2009}
S.~Son and N.~C. Goulbourne, ``{Finite deformations of tubular dielectric
  elastomer sensors},'' {\em Journal of Intelligent Material Systems and
  Structures}, vol.~20, pp.~2187--2199, 2009.

\bibitem{Kofod2001}
G.~Kofod, {\em {Dielectric elastomer actuators}}.
\newblock PhD thesis, Technical University of Denmark, 2001.

\bibitem{Sommer-Larsen2002}
P.~Sommer-Larsen, G.~Kofod, M.~Shridhar, M.~Benslimane, and P.~Gravesen,
  ``{Performance of dielectric elastomer actuators and materials},'' {\em
  Proceedings of SPIE - Smart Structures and Materials}, vol.~4695,
  pp.~158--166, 2002.

\bibitem{Goulbourne2003}
N.~Goulbourne, M.~I. Frecher, E.~M. Mockensturm, and A.~J. Snyder, ``{Modelling
  of a dielectric elastomer diaphragm for a prosthetic blood pump},'' {\em
  Proceedings of SPIE - Smart Structures and Materials}, vol.~5051,
  pp.~319--331, 2003.

\bibitem{Yang2004}
E.~Yang, M.~I. Frecher, E.~M. Mockensturm, and D.~Wu, ``{Analystical model and
  experimental characterization of a dielectric elastomer annulus actuator
  undergoing large quasi-static deformation},'' {\em Proceedings of SPIE -
  Smart Structures and Materials}, vol.~5390, pp.~183--193, 2004.

\bibitem{Yang2006}
E.~Yang, M.~I. Frecher, and E.~M. Mockensturm, ``{Finite element and
  experimental analyses of non-axisymmetric dielectric elastomer actuators},''
  {\em Proceedings of SPIE - Smart Structures and Materials}, vol.~6168,
  pp.~127--135, 2006.

\bibitem{Rosset2008}
S.~Rosset, M.~Niklaus, P.~Dubois, and H.~R. Shea, ``{Mechanical
  characterization of a dielectric elastomer microactuator with ion-implanted
  electrodes},'' {\em Sensors and Actuators A}, vol.~144, pp.~185--193, 2008.

\bibitem{Voltairas2003}
P.~A. Voltairas, D.~I. Fotiadis, and C.~V. Massalas, ``{A theoretical study of
  the hyperelasticity of electro-gels},'' {\em Proceedings of the Royal Society
  A}, vol.~459, pp.~2121--2130, 2003.

\bibitem{Dorfmann2005}
A.~Dorfmann and R.~W. Ogden, ``{Nonlinear electroelasticity},'' {\em Acta
  Mechanica}, vol.~174, no.~3-4, pp.~167--183, 2005.

\bibitem{Mueller2010}
R.~M\"{u}ller, B.~X. Xu, D.~Gross, M.~Lyschik, D.~Schrade, and S.~Klinkel,
  ``{Deformable dielectrics - optimization of heterogeneities},'' {\em
  International Journal of Engineering Science}, vol.~48, pp.~647--657, 2010.

\bibitem{Zwecker2011}
S.~Zwecker, R.~M\"{u}ller, and S.~Klinkel, ``{Nonlinear finite element
  simulation of thin dielectric elastomer structures},'' in {\em Proceedings of
  1st Young Researcher Symposium by Center for Mathematical and COmputational
  Modelling, Kaiserslautern, Germany}, 2011.

\bibitem{Vertechy2012}
R.~Vertechy, G.~Berselli, V.~P. Castelli, and M.~Bergamasco, ``{Continuum
  thermo-electro-mechanical model for electrostrictive elastomers},'' {\em
  Journal of Intelligent Material Systems and Structures}, vol.~24,
  pp.~761--778, Aug. 2012.

\bibitem{Ask2012}
A.~Ask, A.~Menzel, and M.~Ristinmaa, ``{Electrostriction in
  electro-viscoelastic polymers},'' {\em Mechanics of Materials}, vol.~50,
  pp.~9--21, July 2012.

\bibitem{Bueschel2013}
A.~B\"{u}schel, S.~Klinkel, and W.~Wagner, ``{Dielectric elastomers - numerical
  modelling of nonlinear visco-electroelasticity},'' {\em International Journal
  of Numerial Methods in Engineering}, vol.~93, pp.~834--856, 2013.

\bibitem{Simo1987}
J.~C. Simo, ``{On a fully three-dimensional finite-strain viscoelastic damage
  model: Formulation and computational aspects},'' {\em Computer Methods in
  Applied Mechanics and Engineering}, vol.~60, no.~2, pp.~153--173, 1987.

\bibitem{Lion1997}
A.~Lion, ``{A physically based method to represent the thermo-mechanical
  behaviour of elastomers},'' {\em Acta Mechanica}, vol.~123, pp.~1--25, Mar.
  1997.

\bibitem{Lubliner1985}
J.~Lubliner, ``{A model of rubber viscoelasticity},'' {\em Mechanics Research
  Communications}, vol.~12, no.~2, pp.~93--99, 1985.

\bibitem{Reese1998}
S.~Reese and S.~Govindjee, ``{A theory of finite viscoelasticity and numerical
  aspects},'' {\em International Journal of Solids and Structures}, vol.~35,
  no.~26-27, pp.~3455--3482, 1998.

\bibitem{Huber2000}
N.~Huber and C.~Tsakmakis, ``{Finite deformation viscoelasticity laws},'' {\em
  Mechanics of Materials}, vol.~32, pp.~1--18, Jan. 2000.

\bibitem{Saxena2013a}
P.~Saxena, M.~Hossain, and P.~Steinmann, ``{A theory of finite deformation
  magneto-viscoelasticity},'' {\em International Journal of Solids and
  Structures}, vol.~50, no.~24, pp.~3886--3897, 2013.

\bibitem{McMeeking2005}
R.~M. McMeeking and C.~M. Landis, ``{Electrostatic forces and stored energy for
  deformable dielectric materials},'' {\em Journal of Applied Mechanics},
  vol.~72, no.~4, pp.~581--590, 2005.

\bibitem{Coleman1963}
B.~D. Coleman and W.~Noll, ``{The thermodynamics of elastic materials with heat
  conduction and viscosity},'' {\em Archive for Rational Mechanics and
  Analysis}, vol.~13, no.~1, pp.~167--178, 1963.

\bibitem{Holzapfel1996a}
G.~A. Holzapfel and J.~C. Simo, ``{A new viscoelastic constitutive model for
  continuous media at finite thermomechanical changes},'' {\em International
  Journal of Solids and Structures}, vol.~33, no.~20-22, pp.~3019--3034, 1996.

\bibitem{Koprowski-Theiss2011}
N.~Koprowski-Theiss, M.~Johlitz, and S.~Diebels, ``{Characterizing the time
  dependence of filled EPDM},'' {\em Rubber Chemistry and Technology}, vol.~84,
  no.~2, pp.~147--165, 2011.

\bibitem{Dorfmann2006}
A.~Dorfmann and R.~W. Ogden, ``{Nonlinear electroelastic deformations},'' {\em
  Journal of Elasticity}, vol.~82, no.~2, pp.~99--127, 2006.

\end{thebibliography}
\end{document}